\documentclass{article}

 \usepackage[vmargin=1.2in,hmargin=1.1in]{geometry}
\usepackage{amsmath,amssymb,amsfonts,graphicx,amsthm,nicefrac,mathtools,bm}
\usepackage{hyperref}
\usepackage[numbers]{natbib}
\usepackage[english]{babel}
\usepackage{times}
\usepackage[T1]{fontenc}
\usepackage{bbm,mdframed}
\usepackage[dvipsnames]{xcolor}
\usepackage{xspace}
\usepackage{rotating,multirow}
\usepackage{enumerate,graphics,enumitem}

\newtheorem{theorem}{Theorem}

\definecolor{verylightblue}{rgb}{0.7,0.8,1}
  {\begin{mdframed}[backgroundcolor=verylightblue]\begin{theorem}}%
  {\end{theorem}\end{mdframed}}

\definecolor{verylightgray}{gray}{0.95}
  {\begin{mdframed}[backgroundcolor=verylightgray]\begin{proof}}%
  {\end{proof}\end{mdframed}}

\newtheorem{lemma}{Lemma}
\definecolor{verylightred}{rgb}{1,0.8,0.8}
  {\begin{mdframed}[backgroundcolor=verylightred]\begin{lemma}}%
  {\end{lemma}\end{mdframed}}

\newtheorem{proposition}{Proposition}
  {\begin{mdframed}[backgroundcolor=verylightblue]\begin{proposition}}%
  {\end{proposition}\end{mdframed}}

\theoremstyle{definition}

\theoremstyle{remark}

\makeatletter

\newtheorem*{rep@theorem}{\rep@title}
\newcommand{\newreptheorem}[2]
{\newenvironment{rep#1}[1]
{\def\rep@title{#2 \ref{##1}} \begin{rep@theorem}}%
 {\end{rep@theorem}}}
\makeatother
\newreptheorem{theorem}{Theorem}
\newreptheorem{lemma}{Lemma}
\newreptheorem{corollary}{Corollary}
\newreptheorem{proposition}{Proposition}

\newcommand{\figref}[1]{Figure~\ref{fig:#1}}

\newcommand{\secref}[1]{Section~\ref{sec:#1}}
\newcommand{\appref}[1]{Appendix~\ref{app:#1}}

\newcommand{\lemref}[1]{Lemma~\ref{lem:#1}}

\newcommand{\thmref}[1]{Theorem~\ref{thm:#1}}

\newcommand{\tabref}[1]{Table~\ref{tab:#1}}

\newcommand{\PP}[1]{\textnormal{Pr}\!\left\{{#1}\right\}} 
\newcommand{\EE}[1]{\mathbb{E}\left[{#1}\right]} 

\newcommand{\EEst}[2]{\mathbb{E}\left[{#1}\ \middle| \ {#2}\right]} 
\newcommand{\PPst}[2]{\text{Pr}\!\left\{{#1}\ \middle| \ {#2}\right\}} 

\def\R{\mathbb{R}}

\newcommand{\ignore}[1]{}

\newcommand{\eps}{\epsilon}

\usepackage{tikz}
\usetikzlibrary{arrows}
\usetikzlibrary{shapes}

\usepackage{fancyhdr}
\newcommand{\thedate}{\today}
\newcommand{\theauthor}{}
\newcommand{\thetitle}{Online control of the false discovery rate with decaying memory}

\date{\thedate}
\author{\theauthor}
\title{\thetitle}

\usepackage[mmddyy]{datetime}

\newcommand{\nonnulls}{\mathcal{H}^1}
\newcommand{\nulls}{\mathcal{H}^0}

\newcommand{\fdp}{\textnormal{FDP}}
\newcommand{\fdr}{\textnormal{FDR}}
\newcommand{\fwer}{\textnormal{FWER}}
\newcommand{\mfdr}{\textnormal{mFDR}}
\newcommand{\sfdr}{\textnormal{sFDR}}
\newcommand{\memfdr}{\textnormal{mem-FDR}}
\newcommand{\mempower}{\textnormal{mem-power}}
\newcommand{\power}{\textnormal{power}}

\newcommand{\fdrstar}{\fdr_*}
\newcommand{\fdphat}{\widehat{\fdp}}

\newcommand{\One}[1]{{\bf{1}}\left\{{#1}\right\}}

\newcommand{\decay}{\delta}
\newcommand{\Vd}{V^\decay}
\newcommand{\Rd}{R^\decay}
\newcommand{\tVd}{U^\decay}
\newcommand{\tRd}{D^\decay}

\def\N{\mathbb N}

\def\F{\mathcal{F}}

\makeatletter
\newcommand{\dotfrac}[2]{
\mathchoice
{\ooalign{$\genfrac{}{}{0pt}{0}{#1}{#2}$\cr\leavevmode\cleaders\hb@xt@ .22em{\hss $\displaystyle\cdot$\hss}\hfill\kern\z@\cr}}
{\ooalign{$\genfrac{}{}{0pt}{1}{#1}{#2}$\cr\leavevmode\cleaders\hb@xt@ .22em{\hss $\textstyle\cdot$\hss}\hfill\kern\z@\cr}}
{\ooalign{$\genfrac{}{}{0pt}{2}{#1}{#2}$\cr\leavevmode\cleaders\hb@xt@ .22em{\hss $\scriptstyle\cdot$\hss}\hfill\kern\z@\cr}}
{\ooalign{$\genfrac{}{}{0pt}{3}{#1}{#2}$\cr\leavevmode\cleaders\hb@xt@ .22em{\hss $\scriptscriptstyle\cdot$\hss}\hfill\kern\z@\cr}}
}
\makeatother

\usepackage{algorithm,algorithmic}

\newcommand{\defn}{\ensuremath{:\, =}}

\begin{document}

\author{
Aaditya Ramdas ~ ~ Fanny Yang  ~ ~ Martin J. Wainwright  ~ ~  Michael I. Jordan\\
University of California, Berkeley\\
\texttt{$\{$aramdas, fanny-yang, wainwrig, jordan$\}$ @berkeley.edu}
}
\maketitle

\begin{abstract}
In the online multiple testing problem, $p$-values corresponding to
different null hypotheses are observed one by one, and the decision
of whether or not to reject the current hypothesis must be made
immediately, after which the next $p$-value is observed.
Alpha-investing algorithms to control the false discovery rate (FDR),
formulated by Foster and Stine, have been generalized and applied to
many settings, including quality-preserving databases in science and
multiple A/B or multi-armed bandit tests for internet commerce.  This
paper improves the class of generalized alpha-investing algorithms
(GAI) in four ways: (a) we show how to uniformly improve the power of
the entire class of monotone GAI procedures by
awarding more alpha-wealth for each rejection, giving a win-win
resolution to a recent dilemma raised by Javanmard and Montanari, (b)
we demonstrate how to incorporate prior weights to indicate domain
knowledge of which hypotheses are likely to be non-null, (c)
we allow for differing penalties for false discoveries to indicate
that some hypotheses may be more important than others, (d) we define
a new quantity called the decaying memory false discovery rate
($\memfdr$) that may be more meaningful for truly temporal applications, 
and which alleviates problems that we describe and refer to as ``piggybacking'' and
``alpha-death''. Our GAI++ algorithms incorporate all four
generalizations simultaneously, and reduce to more powerful variants
of earlier algorithms when the weights and decay are all set to unity.
Finally, we also describe a simple method to derive new online FDR rules based on an
estimated false discovery proportion.
 \end{abstract}


\section{Introduction}
\label{sec:intro}

The problem of multiple comparisons was first recognized in the
seminal monograph by Tukey~\cite{tukey1953problem}: simply stated,
given a collection of multiple hypotheses to be tested, the goal is to
distinguish between the nulls and non-nulls, with suitable control on
different types of error. We are given access to one $p$-value for each
hypothesis, which we use to decide which subset of hypotheses to
reject, effectively proclaiming the rejected hypothesis as being
non-null. The rejected hypotheses are called \emph{discoveries}, and
the subset of these that were truly null---and hence mistakenly
rejected---are called \emph{false discoveries}.  In this work, we
measure a method's performance using the \emph{false discovery rate}
(FDR)~\cite{BH95}, defined as the expected ratio of false discoveries
to total discoveries. Specifically, we require that any
procedure must guarantee that the FDR is bounded by a pre-specified
constant $\alpha$.

The traditional form of multiple testing is \emph{offline} in nature,
meaning that an algorithm testing $N$ hypotheses receives the entire
batch of $p$-values $\{P_1,\dots,P_N\}$ at one time instant. In the
\emph{online} version of the problem, we do not know how many
hypotheses we are testing in advance; instead, a possibly infinite
sequence of $p$-values appear one by one, and a decision
about rejecting the null must be made before the next $p$-value is
received. There are at least two different motivating justifications for
considering the online setting:
\begin{enumerate}

  \item[M1.] We may have the entire batch of $p$-values available at our
    disposal from the outset, but we may nevertheless choose to
    process the $p$-values one by one in a particular order. Indeed, if
    one can use prior knowledge to ensure that non-nulls typically
    appear earlier in the ordering, then carefully designed online
    procedures could result in more discoveries than offline
    algorithms (that operate without prior knowledge) such as the
    classical Benjamini-Hochberg algorithm~\cite{BH95}, while having
    the same guarantee on FDR control. This motivation underlies one
    of the original online multiple testing paper, namely that
    of~\citet{foster2008alpha}.
  \item[M2.] We may genuinely conduct a sequence of 
    tests one by one, where both the choice of the next null
    hypothesis and the level at which it is tested may depend on the
    results of the previous tests. Motivating applications include the
    desire to provide anytime guarantees for (i) internet companies
    running a sequence of A/B tests over
    time~\cite{javanmard2016online}, (ii) pharmaceutical companies
    conducting a sequence of clinical trials using multi-armed
    bandits~\cite{yang2017multi}, or (iii) quality-preserving
    databases in which different research teams test different
    hypotheses on the same data over
    time~\cite{aharoni2014generalized}.
\end{enumerate}
The algorithms developed in this paper apply to both settings, with
emphasis on motivation M2.

Let us first reiterate the need for corrections when testing a sequence of hypotheses
 in the online setting, even when all the p-values are independent.  
If each hypothesis $i$ is
tested independently of the total number of tests either performed
before it or to be performed after it, then we have no control over
the number of false discoveries made over time. Indeed, if our test
for every $P_i$ takes the form $\One{P_i \leq \alpha}$ for some fixed
$\alpha$, then, while the type $1$ error for any individual test is
bounded by $\alpha$, the set of discoveries could have arbitrarily
poor FDR control. For example, under the ``global null'' where every
hypothesis is truly null, as long as the number of tests $N$ is large
and the null $p$-values are uniform, this method will make at least one
rejection with high probability (w.h.p.), and since in this setting every discovery is
a false discovery, w.h.p. the FDR will equal one.

A natural alternative that takes multiplicity into account is the
Bonferroni correction. If one knew the total number $N$ of tests to be
performed, the decision rule $\One{P_i \leq \alpha/N}$ for each $i \in
\{1,\dots,N\}$ controls the probability of even a single false
discovery---a quantity known as the familywise error rate or FWER---at
level $\alpha$, as can be seen by applying the union bound. The
natural extension of this solution to having an unknown and
potentially infinite number of tests is called \emph{alpha-spending.}
Specifically, we choose any sequence of constants $\{\alpha_i\}_{i \in
  \N}$ such that $\sum_i \alpha_i \leq \alpha$, and on receiving
$P_i$, our decision is simply $\One{P_i \leq \alpha_i}$.  However,
such methods typically make very few discoveries---meaning that they
have very low power---when the number of tests is
large, because they must divide their error budget of $\alpha$, also
called alpha-wealth, among a large number of tests.

Since the FDR is less stringent than FWER, procedures that guarantee
FDR control are generally more powerful, and often far more powerful,
than those controlling FWER.  This fact has led to the wide adoption
of FDR as a de-facto standard for offline multiple testing (note,
e.g., that the Benjamini-Hochberg paper~\cite{BH95} currently has over
40,000 citations).

\citet{foster2008alpha} designed the first online alpha-investing
procedures that use and earn alpha-wealth in order to control a
modified definition of FDR. \citet{aharoni2014generalized} further
extended this to a class of generalized alpha-investing (GAI)
methods, but once more for the modifed FDR. It was only recently that
\citet{javanmard2016online} demonstrated that monotone GAI algorithms,
appropriately parameterized, can control the
(unmodified) FDR for independent $p$-values. It is this last work that
our paper directly improves upon and generalizes; however, as we
summarize below, many of our modifications and generalizations are
immediately applicable to all previous algorithms.

\paragraph{Contributions and outline.} Instead of presenting the most
general and improved algorithms immediately, we choose to present
results in a bottom-up fashion, introducing one new concept at a time
so as to lighten the symbolic load on the reader. For this purpose, we
set up the problem formally in~\secref{setup}.  Our contributions are
organized as follows:

\begin{enumerate}
\item \textbf{Power.}  In \secref{GAI}, we introduce the generalized
  alpha-investing (GAI) procedures, and demonstrate how to uniformly
  improve the power of monotone GAI procedures that control FDR for
  independent $p$-values, resulting in a win-win resolution to a dilemma
  posed by Javanmard and Montanari~\cite{javanmard2016online}. This
  improvement is achieved by a somewhat subtle modification that
  allows the algorithm to reward more alpha-wealth at every rejection
  but the first.  We refer to our algorithms as \emph{improved
    generalized alpha-investing} (GAI++) procedures, and provide
  intuition for why they work through a general super-uniformity lemma
  (see~\lemref{power} in~\secref{lemma}). We also provide
  an alternate way of deriving online FDR procedures by defining and bounding
   a natural estimator for the false discovery proportion $\fdphat$.
\item \textbf{Weights.}  In \secref{weights}, we demonstrate how to
  incorporate certain types of prior information about the different
  hypotheses. For example, we may have a prior weight for each
  hypothesis, indicating whether it is more or less likely to be null.
  Additionally, we may have a
  different penalty weight for each hypothesis, indicating differing
  importance of hypotheses.  
  These prior and penalty weights have been
  incorporated successfully into offline
  procedures~\cite{BH97,genovese2006false,ramdas2017unified}.  In the online setting,
  however, there are some technical challenges that prevent immediate
  application of these offline procedures.  For example, in the
  offline setting all the weights are constants, but in the online
  setting, we allow them to be random variables that depend on the
  sequence of past rejections. Further, in the offline setting all
  provided weights are renormalized to have an empirical mean of one,
  but in the truly online setting (motivation M2) we do not know the sequence of hypotheses
  or their random weights in advance, and hence we cannot perform any
  such renormalization. We clearly outline and handle such issues and
  design novel prior- and/or penalty-weighted GAI++ algorithms that
  control the penalty-weighted $\fdr$ at any time. This may be seen as
  an online analog of doubly-weighted procedures for the offline
  setting~\cite{blanchard2008two,ramdas2017unified}. Setting
  the weights to unity recovers the original class of GAI++
  procedures.

\item \textbf{Decaying memory.} In \secref{decay}, we discuss some
  implications of the fact that existing algorithms have an infinite
  memory and treat all past rejections equally, no matter when they
  occurred.  This causes phenomena that we term as ``piggybacking'' (a
  string of bad decisions, riding on past earned alpha-wealth) and
  ``alpha-death'' (a permanent end to decision-making when the
  alpha-wealth is essentially zero). These phenomena may be desirable
  or acceptable under motivation M1 when dealing with batch problems,
  but are generally undesirable under motivation M2.  To address these
  issues, we propose a new error metric called the \emph{decaying
    memory false discovery rate}, abbreviated as $\memfdr$, that we
  view as better suited to multiple testing for truly temporal
  problems. Briefly, $\memfdr$ pays more attention to recent
  discoveries by introducing a user-defined discount factor, $0 <
  \decay \leq 1$, into the definition of FDR. We demonstrate how to
  design GAI++ procedures that control online $\memfdr$, and show that
  they have a stable and robust behavior over time. Using $\decay < 1$
  allows these procedures to slowly forget their past decisions
  (reducing piggybacking), or they can temporarily ``abstain'' from
  decision-making (allowing rebirth after alpha-death).  Instantiating
  $\decay = 1$ recovers the class of GAI++ procedures.
\end{enumerate}

We note that the generalizations to incorporate weights and decaying
memory are entirely orthogonal to the improvements that we introduce
to yield GAI++ procedures, and hence these ideas immediately extend to
other GAI procedures for non-independent $p$-values. We also describe
simulations involving several of the aforementioned generalizations
in~\appref{sims}.


\section{Problem Setup}\label{sec:setup}

At time $t=0$, before the $p$-values begin to appear, we fix the level
$\alpha$ at which we wish to control the FDR over time. At each time
step $t = 1, 2,\dots$, we observe a $p$-value $P_t$ corresponding to
some null hypothesis $H_t$, and we must immediately decide whether to
reject $H_t$ or not. If the null hypothesis is true, $p$-values are
stochastically larger than the uniform distribution
(``super-uniform'', for short), formulated as follows: if $\nulls$ is
the set of true null hypotheses, then for any null $H_t\in\nulls$, we
have
\begin{align}
\label{eqn:SuperUniform}
\PP{P_t \leq x} \leq x \quad\mbox{for any $x \in [0,1]$.}
\end{align}
We do not make assumptions on the marginal distribution of the $p$-values
for hypotheses that are non-null / false.
Although they can be arbitrary, it is useful to think of them as being
stochastically smaller than the uniform distribution, since only then do
they carry signal that differentiates them from nulls.
%
Our task is to design threshold levels $\alpha_t$ according to which
we define the rejection decision as $R_t = \One{P_t \leq \alpha_t}$,
where $\One{\cdot}$ is the indicator function.  Since the aim is to
control the $\fdr$ at the fixed level $\alpha$ at any time $t$, each
$\alpha_t$ must be set according to the past decisions of the
algorithm, meaning that $\alpha_t = \alpha_t(R_1,\dots,R_{t-1})$.
Note that, in accordance with past work, we require that $\alpha_t$
does not directly depend on the observed $p$-values but only on past
rejections. Formally, we define the sigma-field at time $t$ as $\F^t =
\sigma(R_1,\dots,R_t)$, and insist that
\begin{align}
 \alpha_t \in \F^{t-1} ~~\equiv~~ \text{$\alpha_t$ is
   $\F^{t-1}$-measurable} ~~\equiv~~ \text{$\alpha_t$ is
   \emph{predictable}.}
\end{align}
 As studied by~\citet{javanmard2015online}, and as is predominantly
 the case in offline multiple testing, we consider \emph{monotone}
 decision rules, where $\alpha_t$ is a coordinate-wise nondecreasing
 function:
 \begin{equation}
   \label{eqn:monotone}
 \text{ if } \tilde R_i \geq R_i \text{ for all } i \leq t-1 , \text{
   then we have } \alpha_t(\tilde R_1,\dots,\tilde R_{t-1}) \geq
 \alpha_t(R_1,\dots,R_{t-1}).
  \end{equation}
Existing online multiple testing algorithms control some variant of the FDR over
time, as we now define. At any time $T$, let $R(T) = \sum_{t=1}^T R_t$
be the total number of rejections/discoveries made by the algorithm so
far, and let $V(T) = \sum_{t \in \nulls} R_t$ be the number of false
rejections/discoveries. Then, the false discovery proportion and rate are defined as
\begin{align*}
 \fdp(T) := \dotfrac{ V(T) }{R(T)} \text{~ and ~} \fdr(T) = \EE{ \dotfrac{ V(T) }{R(T)} } ,
\end{align*}
where we use the dotted-fraction notation corresponds to the shorthand
$\dotfrac{a}{b} = \frac{a}{b \vee 1}$. Two variants of the FDR studied
in earlier online FDR works \cite{foster2008alpha,javanmard2015online}
are the \emph{marginal} FDR given by \mbox{$\mfdr_\eta(T) = \frac{
    \EE{ V(T) }}{ \EE{R(T)} + \eta}$,} with a special case being
\mbox{$\mfdr(T) = \frac{\EE{ V(T) }}{ \EE{R(T)\vee 1} }$,} and the
\emph{smoothed} FDR, given by \mbox{$\sfdr_\eta(T) = \EE{ \frac{ V(T)
    }{R(T) + \eta} }.$}
In \appref{previous}, we summarize a
variety of algorithms and dependence assumptions considered in previous work.


\subsection{Summary of joint dependence assumptions in previous work}\label{app:previous}

We use the phrase ``$\fdrstar$ control'' to mean the control of either
$\fdr$ or $\mfdr$ or $\sfdr$. 
It is important to discuss the assumptions on the joint dependence on $p$-values, under which $\fdrstar$ control can be proved. These are listed from (approximately) weakest to strongest below:

\begin{enumerate}
\item \textbf{Arbitrary Dependence.} Null $p$-values are arbitrarily dependent on all other $p$-values.
\item \textbf{SuperCoRD.} Null $p$-values are super-uniform conditional on the time of most recent discovery, meaning that for all $t \in \nulls$ and for any $\alpha_t \in \F^{t-1}$, we have
\[
\PPst{P_t \leq \alpha_t}{\tau_{\text{prev}}} \leq \alpha_t,
\]
where $\tau_{\text{prev}} = \max_{s < t} \{s : R_s = 1\}$ is the time of the previous rejection.
\item \textbf{SuperCoND.} Null $p$-values are super-uniform conditional on the number of discoveries up to that point, meaning that for all $t \in \nulls$ and for any $\alpha_t \in \F^{t-1}$, we have
\[
\PPst{P_t \leq \alpha_t}{R(T-1)} \leq \alpha_t.
\]
\item \textbf{SuperCoAD.} Null $p$-values are super-uniform conditional on  all discoveries, meaning that for all $t \in \nulls$ and for any $\alpha_t \in \F^{t-1}$, we have
\[
\PPst{P_t \leq \alpha_t}{\F^{t-1}} \leq \alpha_t.
\]
\item \textbf{Independence.} Null $p$-values are independent of all other $p$-values.
\end{enumerate}

Table~\ref{tab:summary} summarizes some known algorithms, the dependence these algorithms can handle, and the type of FDR control they guarantee. Of special note is an algorithm called LORD~\cite{javanmard2016online} that the authors noted performs consistently well in practice, and thus will be the focus of most of our experiments (the conclusions of which carry forward qualitatively to other monotone algorithms).

\begin{table}[h!]
\small
\centering
\begin{tabular}{ | l l l l l |}
\hline 
Ref. & Algorithm  &  Dependence  &  Control (at any $T$) & Monotone? \\
 \hline 
- & Alpha-spending & Arbitrary &  $\fwer(T)$ & No \\
\cite{foster2008alpha} & Alpha-investing (AI) & SuperCoAD &  $\mfdr_\eta(T)$ & No \\
\cite{aharoni2014generalized} & Generalized Alpha-investing (GAI) & SuperCoAD & $\mfdr_\eta(T)$ & No \\
\cite{javanmard2015online} & Levels based on Number of Discoveries (LOND) & SuperCoND &  $\mfdr_\eta(T), \fdr(T)$ & Yes \\
\cite{javanmard2015online} & LOND (with a conservative correction) & Arbitrary & $\fdr(T)$ & Yes \\
\cite{javanmard2015online} & Levels based on most Recent Disc. (LORD'15) & SuperCoRD & $\mfdr_\eta(T)$
& Yes \\
\cite{javanmard2016online} & Monotone GAI (including LORD'17) & Independence & $\fdr(T), \sfdr_\eta(T)$ & Yes\\
\hline
\end{tabular}
\caption{Summary of previous work. 
Note that LORD'17 is an improvement over LORD'15 with higher power, and the shorthand ``LORD'' will be reserved for LORD'17.} \label{tab:summary}
\end{table}

\section{Generalized alpha-investing (GAI) rules}
\label{sec:GAI}

The generalized class of alpha-investing
rules~\cite{aharoni2014generalized} essentially covers most rules that
have been proposed thus far, and includes a wide range of algorithms
with different behaviors.  In this section, we present a uniform
improvement to monotone GAI algorithms for FDR control under
independence.

Any algorithm of the GAI type begins with an \emph{alpha-wealth} of
$W(0)=W_0 > 0$, and keeps track of the wealth $W(t)$ available after
$t$ steps. At any time $t$, a part of this alpha-wealth is used to
test the $t$-th hypothesis at level $\alpha_t$, and the wealth is
immediately decreased by an amount $\phi_t$. If the $t$-th hypothesis
is rejected, that is if $R_t := \One{P_t \leq \alpha_t} = 1$, then we
award extra wealth equaling an amount $\psi_t$. Recalling the
definition $\F^t \defn \sigma(R_1,\dots,R_t)$, we require that
$\alpha_t, \phi_t, \psi_t \in \F^{t-1}$, meaning they are predictable,
and $W(t) \in \F^t$, with the explicit update $W(t) \defn W(t-1) -
\phi_t + R_t \psi_t$. The parameters $W_0$ and the sequences
$\alpha_t,\phi_t,\psi_t$ are all user-defined. They must be chosen so
that the total wealth $W(t)$ is always non-negative, and hence that
$\phi_t \leq W(t-1)$ If the wealth ever equals zero, the procedure is
not allowed to reject any more hypotheses since it has to choose
$\alpha_t$ equal to zero from then on.
The only real restriction for $\alpha_t,\phi_t,\psi_t$ arises
from the goal to control $\fdr$.
This condition takes a natural form---whenever a rejection takes place,
we cannot be allowed to award an arbitrary amount of wealth. Formally,
for some user-defined constant $B_0$, we must have
\begin{equation}\label{eqn:GAIreward}
  \psi_t \leq \min\{ \phi_t + B_0, \frac{\phi_t}{\alpha_t} + B_0 - 1 \}.
\end{equation}
Many GAI rules are not monotone (cf. equation~\eqref{eqn:monotone}),
meaning that $\alpha_t$ is not always a coordinatewise nondecreasing
function of $R_1,\dots,R_{t-1}$, as mentioned in the last column of
\tabref{summary} (\appref{previous}).  \tabref{examples} has some
examples, where $\tau_k := \min_{s \in \N} \One{\sum_{t=1}^s R_t = k}$
is the time of the $k$-th rejection.

\begin{table}[h]
\small \centering
\begin{tabular}{ | l l l l l |}
\hline 
Name & Parameters & Level $\alpha_t$  &  Penalty $\phi_t$ & Reward $\psi_t$  \\ 
\hline 
\cite{foster2008alpha} Alpha-investing (AI) & --- & $\frac{\phi_t}{1+\phi_t}$ & $\leq W(t-1)$ & $\phi_t + B_0$  \\ 
\cite{aharoni2014generalized} Alpha-spending with rewards  & $\kappa \leq 1,c$  & $c W(t-1)$ & $\kappa W(t-1)$ & satisfy \eqref{eqn:GAIreward} \\ 
\cite{javanmard2016online} LORD'17 & $\sum\limits_{i=1}^\infty \gamma_i = 1$ & $\phi_t$ & $\gamma_t W_0 + B_0\sum\limits_{j: \tau_j < t}\gamma_{t-\tau_j} $ & $B_0=\alpha-W_0$\\ 
\hline
\end{tabular}
\caption{Examples of GAI rules. } \label{tab:examples}
\end{table}
%




\subsection{Improved monotone GAI rules (GAI++) under independence}

In their initial work on GAI rules, \citet{aharoni2014generalized} did
not incorporate an explicit parameter $B_0$; rather, they proved that
choosing $W_0=B_0=\alpha$ suffices for $\mfdr_1$ control.
 In subsequent work, \citet{javanmard2016online} introduced
the parameter $B_0$ and proved that for monotone GAI rules, the same
choice $W_0 = B_0 = \alpha$ suffices for $\sfdr_1$ control, whereas
the choice $B_0 = \alpha - W_0$ suffices for FDR control, with both
results holding under independence. In fact, their monotone GAI rules
with $B_0 = \alpha - W_0$ are the only known methods that control FDR.
This state of affairs leads to the following dilemma raised in their
paper~\cite{javanmard2016online}:
\begin{quote}
{\small A natural question is whether, in practice, we should choose
  $W_0, B_0$ as to guarantee FDR control (and hence set $B_0 = \alpha
  - W_0 \ll \alpha$) or instead be satisfied with $\mfdr$ or $\sfdr$
  control, which allow for $B_0 = \alpha$ and hence potentially larger
  statistical power.}
\end{quote}

Our first contribution is a ``win-win'' resolution to this dilemma:
more precisely, we prove that we can choose $B_0 = \alpha$ while
maintaining FDR control, with a small catch that at the \emph{very
  first rejection only}, we need $B_0 = \alpha - W_0$. Of course, in
this case $B_0$ is not constant, and hence we replace it by the random
variable $b_t \in \F^{t-1}$, and we prove that choosing $W_0,b_t$ such
that $b_t + W_0 = \alpha$ for the first rejection, and simply $b_t =
\alpha$ for every future rejection, suffices for formally proving FDR
control under independence. This achieves the best of both worlds
(guaranteeing FDR control, and handing out the largest possible reward
of $\alpha$), as posed by the above dilemma. To restate our
contribution, we effectively prove that the power of monotone GAI
rules can be uniformly improved without changing the FDR guarantee.

Formally, we define our improved generalized alpha-investing (GAI++)
algorithm as follows. It sets $W(0) = W_0$ with $0 \leq W_0 \leq
\alpha$, and chooses $\alpha_t\in \F^{t-1}$ to make decisions $R_t =
\One{P_t \leq \alpha_t}$ and updates the wealth $W(t) = W(t-1) -
\phi_t + R_t \psi_t \in \F^t$ using some $\phi_t \leq W(t-1) \in
\F^{t-1}$ and some reward $\psi_t \leq \min\{ \phi_t + b_t,
\frac{\phi_t}{\alpha_t} + b_t - 1 \} \in \F^{t-1}$, using the choice
\begin{align*}
b_t = \begin{cases} \alpha - W_0 \quad \text{ when ~ } R(t-1) = 0
  \\ \alpha \quad \quad \quad \text{ ~ otherwise } \end{cases} \in
\F^{t-1}.
\end{align*}
As an explicit example, given an infinite nonincreasing sequence of positive constants $\{\gamma_j\}$ that sums to one, the LORD++ algorithm effectively makes the choice:
\begin{align}\label{eqn:LORD++}
\alpha_t = \gamma_{t} W_0 + (\alpha - W_0) \gamma_{t -\tau_1} + \alpha \sum_{j: \tau_j < t, \tau_j \neq \tau_1} \gamma_{t-\tau_j},
\end{align}
recalling that $\tau_j$ is the time of the $j$-th rejection. Reasonable default choices include $W_0=\alpha/2$, and $\gamma_j = 0.0722 \frac{\log( j \vee 2)}{j e^{\sqrt {\log j}}}$, the latter derived in the context of testing if a Gaussian is zero mean \cite{javanmard2016online}. 

Any monotone GAI++ rule comes with the following guarantee.

\vspace{0.05in}
\begin{theorem}\label{thm:gai++}
Any monotone GAI++ rule satisfies the bound $ \EE{\dotfrac{V(T) +
    W(T)}{R(T)}} \leq \alpha ~\textnormal{~ for all $T \in \mathbb{N}$}$ under independence. 
Since $W(T) \geq 0$ for all $T \in \N$, any such rule (a)
controls FDR at level $\alpha$ under independence, and (b) has power
at least as large as the corresponding GAI algorithm.
\end{theorem}
The proof of this theorem is provided in~\appref{thmproof++}. Note
that for monotone rules, a larger alpha-wealth reward at each
rejection yields a possibly higher power, but never lower power,
immediately implying statement (b).  Consequently, we provide only a
proof for statement (a) in~\appref{thmproof++}.  For the reader
interested in technical details, a key super-uniformity~\lemref{power}
and associated intuition for online FDR algorithms is provided
in~\secref{lemma}.

\subsection{Intuition for larger rewards via a super-uniformity lemma}
\label{sec:lemma}

For the purposes of providing some intuition for why we are able to
obtain larger rewards than \citet{javanmard2016online}, we present the
following lemma.  In order to set things up, recall that \mbox{$R_t =
  \One{P_t \leq \alpha_t}$} and note that $\alpha_t$ is
$\F^{t-1}$-measurable, being a coordinatewise nondecreasing function
of $R_1,\ldots,R_{t-1}$. Hence, the marginal super-uniformity
assumption~\eqref{eqn:SuperUniform} immediately implies that for
independent $p$-values, we have
\begin{align}
  \label{eqn:superuniformity-cond}
 \PPst{P_t \leq \alpha_t}{\F^{t-1}} \leq \alpha_t, \quad
 \mbox{or equivalently,} \quad \EEst{\dotfrac{\One{P_t \leq
       \alpha_t}}{\alpha_t}}{\F^{t-1}} \leq 1.
\end{align}
Lemma~\ref{lem:power} states that under independence, the above
statement remains valid in much more generality.

Given a sequence $P_1,P_2,\dots$ of independent $p$-values, we first define a
filtration via the sigma-fields of rejection decisions $\F^{i-1} \defn
\sigma(R_1,\ldots,R_{i-1})$, where $R_i \defn \One{P_i \leq
  f_i(R_1,\ldots,R_{i-1})}$ for some coordinatewise nondecreasing
function $f_i: \{0,1\}^{i-1} \to \R$.  With this set-up, we have the
following guarantee:

\begin{lemma} \label{lem:power}
  Let $g: \{0,1\}^T \to \R$ be any coordinatewise nondecreasing
  function such that $g(\vec{x}) > 0$ for any vector $\vec x \neq
  (0,\ldots,0)$. Then for any index $t \leq T$ such that $H_t \in
  \nulls$, we have
\begin{align}
 \EEst{ \dotfrac{\One{P_t \leq
      f_t(R_1,\dots,R_{t-1})}}{g(R_1,\dots,R_T)}}{\mathcal{F}^{t-1}}
\leq \EEst{\dotfrac{f_t(R_1,\dots,R_{t-1})}{g(R_1,\dots,R_T)}}{
  \F^{t-1} }.
\end{align}
\end{lemma}
This super-uniformity lemma is analogous to others used in offline multiple testing~\cite{blanchard2008two,ramdas2017unified}, and will be needed in its full
generality later in the paper. The proof of this lemma in \appref{lemproof} is based on a leave-one-out technique which is common in the multiple testing literature~\cite{heesen2014dynamic,li2016multiple,ramdas2017unified}; ours specifically generalizes a lemma in the Appendix of \citet{javanmard2016online}.

As mentioned, this lemma helps to provide some intuition for the
condition on $\psi_t$ and the unorthodox condition on $b_t$. Indeed,
note that by definition,
\[
\fdr(T) = \EE{\dotfrac{V(T)}{R(T)}} =  \EE{ \dotfrac{ \sum_{t \in \nulls} \One{P_t \leq \alpha_t}}{R(T)} } \leq \EE{ \dotfrac{\sum_{t=1}^T \alpha_t}{\sum_{t=1}^T R_t } } ,
\] 
where we applied \lemref{power} to the coordinatewise nondecreasing
function $g(R_1,\dots,R_T) = R(T)$.  From this equation, we may
infer the following: If $\sum_t R_t = k$, then the FDR will be bounded
by $\alpha$ as long as the total alpha-wealth $\sum_t \alpha_t$ that
was used for testing is smaller than $k \alpha$. In other words, with
every additional rejection that adds one to the denominator, the
algorithm is allowed extra alpha-wealth equaling $\alpha$ for
testing.

In order to see where this shows up in the algorithm design, assume
for a moment that we choose our penalty as $\phi_t = \alpha_t$. Then,
our condition on rewards $\psi_t$ simply reduces to $\psi_t \leq b_t$.
Furthermore, since we choose $b_t = \alpha$ after every rejection
except the first, our total \emph{earned} alpha-wealth is
approximately $\alpha R(T)$, which also upper bounds the total
alpha-wealth used for testing.

The intuitive reason that $b_t$ cannot equal $\alpha$ at the very
first rejection can also be inferred from the above equation. Indeed,
note that because of the definition of FDR, we have $
\dotfrac{V(T)}{R(T)} := \frac{V(T)}{R(T) \vee 1}, $ the denominator
$R(T) \vee 1 = 1$ when the number of rejections equals zero or one.
Therefore, the denominator only starts incrementing at the second
rejection. Hence, the sum of $W_0$ and the first reward must be at
most $\alpha$, following which one may award $\alpha$ at every
rejection. This is the central piece of intuition behind the GAI
algorithm design, its improvement in this paper, and the FDR control
analysis.  To the best of our knowledge, this is the first explicit
presentation for the intuition for online FDR control.

\section{A direct method for deriving new online FDR rules}

Many offline FDR procedures can be derived 
 in terms of an estimate $\fdphat$ of the false discovery proportion; see \citet{ramdas2017unified} and references therein. 
The discussion in \secref{lemma} suggests that it is also possible to write online FDR rules 
in this fashion. Indeed, given any non-negative, predictable sequence $\{\alpha_t\}$, we propose the following definition:
\[
\fdphat(t) \defn \dotfrac{\sum_{j=1}^t \alpha_j}{ R(t) }.
\]
This definition is intuitive because $\fdphat(t)$ approximately overestimates the unknown $\fdp(t)$:
\[
\fdphat(t) \geq \dotfrac{\sum_{j \leq t, j \in \nulls} \alpha_j}{ R(t) } \approx \dotfrac{\sum_{j \leq t, j \in \nulls} \One{P_j \leq \alpha_j}}{ R(t) } = \fdp(t).
\]

 A more direct way to construct new online FDR procedures is to ensure that $\sup_{t \in \N}\fdphat(t)\leq \alpha$, bypassing the use of wealth, penalties and rewards in GAI. This idea is formalized below.

\vspace{0.05in}
\begin{theorem}\label{thm:newview}
For any predictable sequence $\{\alpha_t\}$ such that
$\sup_{t \in \mathbb{N}} \fdphat(t) \leq \alpha$, we have: \\ 
(a) If the p-values are super-uniform
conditional on all past discoveries, meaning that $\PPst{P_j \leq
  \alpha_j}{F^{j-1}} \leq \alpha_j$, then the associated procedure has
$\sup_{T \in \mathbb{N}} \mfdr(T)\leq \alpha$. \\ (b) If the p-values
are independent and if $\{\alpha_t\}$ is monotone, then we also have
$\sup_{T \in \mathbb{N}}\fdr(T) \leq \alpha$.
\end{theorem}

The proof of this theorem is given in \appref{proof-newview}. In our opinion, it is more transparent to verify that LORD++ controls both $\mfdr$ and $\fdr$ using \thmref{newview} than using \thmref{gai++}.

\section{Incorporating prior and penalty weights}\label{sec:weights}

Here, we develop GAI++ algorithms that incorporate
\emph{prior weights} $w_t$, which allow the user to exploit domain knowledge
about which hypotheses are more likely to be non-null, as
well as \emph{penalty weights} $u_t$ to differentiate more
important hypotheses from the rest. 
 The weights must be strictly positive, predictable (meaning that $w_t,u_t \in \F_{t-1}$) and monotone (in the sense of definition \eqref{eqn:monotone}).

\paragraph{Penalty weights. } For many motivating applications, including
internet companies running a series of A/B tests over time, or drug
companies doing a series of clinical trials over time, it is natural
to assume that some tests are more important than others, in the sense
that some false discoveries may have more lasting positive/negative
effects than others. To incorporate this in the offline setting,
\citet{BH97} suggested associating each test with a positive
\emph{penalty} weight $u_i$ with hypothesis $H_i$.  Choosing $u_i > 1$
indicates a more impactful or important test, while $u_i < 1$ means
the opposite.  Although algorithms exist in the offline setting that
can intelligently incorporate penalty weights, no such flexibility
currently exists for online FDR algorithms.  With this motivation in
mind and following \citet{BH97}, define the penalty-weighted FDR as
\begin{align}
  \fdr_u(T) & \defn \EE{\dotfrac{V_u(T)}{R_u(T)}}
\end{align}
where $V_u(T) \defn \sum_{t\in \nulls} u_t R_t = V_u(T-1) + u_T
R_T\One{T \in \nulls}$ and $R_u(T) \defn R_u(T-1) + u_T R_T$.  One may
set $u_t = 1$ to recover the special case of no penalty weights. In the offline
setting, a given set of penalty weights can be rescaled to make the
average penalty weight equal unity, without affecting the associated
procedure. However, in the online setting, we choose penalty weights
$u_t$ one at a time, possibly not knowing the total number of
hypotheses ahead of time.  As a consequence, these weights cannot be
rescaled in advance to keep their average equal to unity. 
 It is important to
note that we allow $u_t \in \F_{t-1}$ to be determined \emph{after} viewing the
past rejections, another important difference from the offline
setting. Indeed, if the hypotheses are logically related (even if the p-values are independent),
then the current hypothesis can be more or less critical depending on which other ones 
are already rejected.

\paragraph{Prior weights. } In many applications, one may have access to
prior knowledge about the underlying state of nature (that is, whether
the hypothesis is truly null or non-null). For example, an older
published biological study might have made significant discoveries, or
an internet company might know the results of past A/B tests or
decisions made by other companies. This knowledge may be incorporated
by a weight $w_t$ which indicates the strength of a prior belief about
whether the hypothesis is null or not---typically, a larger $w_t > 1$
can be interpreted as a greater likelihood of being a non-null,
indicating that the algorithm may be more aggressive in deciding
whether to reject $H_t$.  Such $p$-value weighting was first suggested
in the offline FDR context by~\cite{genovese2006false}, though earlier
work employed it in the context of FWER control. As with penalty
weights in the offline setting, offline prior weights are also usually
rescaled to have unit mean, and then existing offline algorithms
simply replace the $p$-value $P_t$ by the weighted $p$-value
$P_t/w_t$. However, it is not obvious how to incorporate
prior weights in the online setting. As we will see in the sections to come, the online FDR
algorithms we propose will also use $p$-value reweighting; moreover, the
rewards must be prudently adjusted to accommodate the fact that an
a-priori rescaling is not feasible. Furthermore, as opposed to the
offline case, the weights $w_t \in \F_{t-1}$ are allowed to depend on past rejections. This
additional flexibility allows one to set the weights not only based on
our prior knowledge of the current hypothesis being tested, but also
based on properties of the sequence of discoveries (for example,
whether we recently saw a string of rejections or non-rejections). We
point out some practical subtleties with the use and interpretation of
prior weights in~\appref{weights}.


\paragraph{Doubly-weighted GAI++ rules.}  
Given a testing level $\alpha_t$ and weights $w_t, u_t$, all three being predictable and monotone, we make the decision
\begin{align}
  \label{eqn:weighted-decision}
R_t & \defn \One{ P_t \leq \alpha_t u_t w_t}.
\end{align}
This agrees with the intuition that larger prior weights should be
reflected in an increased willingness to reject the null, and we
should favor rejecting more important hypotheses.  As before, our
rejection reward strategy differs before and after $\tau_1$, the time
of the first rejection.  Starting with some \mbox{$W(0) = W_0 \leq
  \alpha$,} we update the wealth as $W(t) = W(t-1) - \phi_t + R_t
\psi_t$, where $w_t,u_t,\alpha_t,\phi_t,\psi_t \in \F^{t-1}$ must be
chosen so that $\phi_t \leq W(t-1)$, and the rejection reward $\psi_t$
must obey the condition
\begin{subequations}
  \begin{align}
    \label{eqn:weighted-reward}
0 \leq \psi_t & \leq \min\left\{ \phi_t + u_t b_t, \frac{\phi_t
 }{u_t w_t \alpha_t} + u_t b_t - u_t \right\}, ~\text{~
  where ~}~\\
\label{eqn:weighted-reward2} b_t &:= \alpha -
\frac{W_0}{u_t}\One{\tau_1 > t-1}\in \F_{t-1}.
\end{align}
\end{subequations}
Notice
that setting $w_t = u_t = 1$ immediately recovers the GAI updates.
Let us provide some intuition for the form of the rewards $\psi_t$,
which involves an interplay between the weights $w_t, u_t$, the
testing levels $\alpha_t$ and the testing penalties $\phi_t$.  First note that large weights $u_t, w_t > 1$ result in a
smaller earning of alpha-wealth and if $\alpha_t, \phi_t$ are
fixed, then the maximum ``common-sense'' weights are determined by
requiring $\psi_t \geq 0$. The requirements of lower rewards for
larger weights and of a maximum allowable weight should both seem
natural; indeed, there must be some price one must pay for an easier
rejection, otherwise we would always use a high prior weight or penalty weight to get
more power, no matter the hypothesis!  We show that such a price does not
have to be paid in terms of the FDR guarantee---we prove that $\fdr_u$
is controlled for any choices of weights---but a
price is paid in terms of power, specifically the ability to make rejections in the
future. Indeed, the combined use of $u_t, w_t$ in both the decision rule $R_t$ and
the earned reward $\psi_t$ keeps us honest; if we overstate our prior belief
in the hypothesis being non-null or its importance by assigning a large $u_t, w_t > 1$, we
will not earn much of a reward (or even a negative reward!), while if
we understate our prior beliefs by assigining a small $u_t, w_t < 1$, then
we may not reject this hypothesis. Hence, it is
prudent to not misuse or overuse the weights, and we recommend that the
scientist uses the default $u_t = w_t = 1$ in practice unless there truly is
prior evidence against the null or a reason to believe the finding would be of importance, 
perhaps due to past studies by other groups or companies, logical relationships between hypotheses, or due to extraneous reasons suggested by the underlying science.


We are now ready to state a theoretical guarantee for the
doubly-weighted GAI++ procedure:

\vspace{0.05in}
\begin{theorem}
  \label{thm:gai++weighted}
Under independence, the doubly-weighted GAI++ algorithm satisfies the
bound $ \EE{\dotfrac{V_u(T) + W(T)}{R_u(T)}} \leq \alpha \textnormal{
  for all $T \in \N$}.  $ Since $W(T) \geq 0$, we also have $\fdr_u(T)
\leq \alpha$ for all $T \in \N$.
\end{theorem}
The proof of this theorem is given in~\appref{thmproofweights}.  
It is important to note that although we provide the proof here only for
GAI++ rules under independence, the ideas would actually carry forward in an
analogous fashion for GAI rules under various other forms of
dependence.


\section{From infinite to decaying memory}
\label{sec:decay}

Here, we summarize two phenomena : (i) the ``piggybacking''
problem that can occur with non-stationary null-proportion, (ii) the
``alpha-death'' problem that can occur with a sequence of nulls. We
propose a new error metric, the decaying-memory FDR ($\memfdr$), that
for truly temporal multiple testing
scenarios, and propose an adjustment of our GAI++ algorithms to
control this quantity.

\paragraph{Piggybacking.} As outlined in motivation M1, when the full
batch of $p$-values is available offline, online FDR algorithms have an
inherent asymmetry in their treatment of different $p$-values, and make
different rejections depending on the order in which they process the
batch. Indeed, \citet{foster2008alpha} demonstrated that if one knew a
reasonably good ordering (with non-nulls arriving earlier), then their
online alpha-investing procedures could attain higher power than the
offline BH procedure. This is partly due to a phenomenon that we call
``piggybacking''---if a lot of rejections are made early, these
algorithms earn and accumulate enough alpha-wealth to reject later hypotheses
more easily by testing them at more lenient thresholds
than earlier ones. 
In essence, later tests
``piggyback'' on the success of earlier tests.  While piggybacking may
be desirable or acceptable under motivation M1, such behavior may be
unwarranted and unwanted under motivation M2. We argue that
piggybacking may lead to a spike in the false discovery rate
\emph{locally in time}, even though the FDR over all time is
controlled.  This may occur when the sequence of hypotheses is
non-stationary and clustered, when strings of nulls may follow strings
of non-nulls.
For concreteness, consider the setting in \citet{javanmard2015online}
where an internet company conducts many A/B tests over
time.  In ``good times'', when a large fraction tests are truly non-null, the company may
accumulate wealth due to frequent rejections. We demonstrate using
simulations that such accumulated wealth can lead to a string of false
discoveries when there is a quick transition to a ``bad period'' where
the proportion of non-nulls is much lower, causing a spike in the false discovery proportion
\emph{locally in time}.

\paragraph{Alpha-death.} Suppose we test a long stretch of nulls,
followed by a stretch of non-nulls.  In this setting, GAI algorithms
will make (almost) no rejections in the first stretch, losing nearly
all of its wealth. Thereafter, the algorithm may be effectively
condemned to have no power, unless a non-null with extremely strong
signal is observed.  Such a situation, from which no recovery is
possible, is perfectly reasonable under motivation M1. The
alpha-wealth has been used up fully, and those are the only rejections
we are allowed to make with that batch of $p$-values. However, for an
internet company operating with motivation M2, it might be
unacceptable to inform them that they essentially cannot run any more
tests, or that they may perhaps never make another useful
discovery.

Both of these problems, demonstrated in simulations
in~\appref{piggyplots}, are due to the fact that the process
effectively has an infinite memory. In the following, we propose one
way to smoothly forget the past and to some extent alleviate the
negative effects of the aforementioned phenomena.


\subsection{Decaying memory false discovery rate ($\memfdr$)}
For a user-defined decay parameter $\delta > 0$, define $\Vd(0) =
\Rd(0) = 0$ and define the \emph{decaying memory FDR} as follows:
\begin{align*}
 \memfdr(T) & \defn \EE{\dotfrac{\Vd(T)}{\Rd(T)}},
\end{align*}
 where $\Vd(T) \defn \decay \Vd(T-1) + R_T\One{T \in \nulls} =
 \sum_{t\in \nulls} \decay^{T-t} R_t\One{t \in \nulls}$, and
 analogously $\Rd(T) \defn \decay \Rd(T-1) + R_T = \sum_{t}
 \decay^{T-t} R_t.  $ This notion of FDR control, which is arguably
 natural for modern temporal applications, appears to be novel in the
 multiple testing literature.  The parameter $\delta$ is reminiscent
 of the discount factor in reinforcement learning.

\paragraph{Penalty-weighted decaying-memory FDR.}  We may naturally
extend the notion of decaying-memory FDR to encompass penalty weights.
Setting $\Vd_u(0)=\Rd_u(0)=0$, we define
\begin{align*}
 \memfdr_u(T) & \defn \EE{\dotfrac{\Vd_u(T)}{\Rd_u(T)}},
\end{align*}
where we define $\Vd_u(T) \defn \decay \Vd_u(T-1) + u_T R_T \One{T \in
  \nulls} = \sum_{t=1}^T \decay^{T-t} u_t R_t \One{t \in \nulls}$,
$\Rd_u(T) \defn \decay \Rd_u(T-1) + u_t R_t = \sum_{t=1}^T
\decay^{T-t} u_t R_t.  $


\paragraph{mem-GAI++ algorithms with decaying memory and weights.}
Given a testing level $\alpha_t$, we make the decision using
equation~\eqref{eqn:weighted-decision} as before, starting with a
 wealth of $W(0) = W_0 \leq \alpha$.  Also, recall that $\tau_k$ is the time of the $k$-th rejection.
On making the decision $R_t$, we update the wealth as:
\begin{align}
W(t) & \defn \decay W(t-1) + (1-\decay)W_0\One{\tau_1 > t - 1} -
\phi_t + R_t \psi_t \label{eqn:wealth-update},\\ 
\text{ so that ~}~ W(T) &= W_0 \decay^{T - \min\{\tau_1, T\}} + \sum_{t=1}^T \decay^{T-t} (-\phi_t + R_t
\psi_t).\nonumber
\end{align}
The first term in equation~\eqref{eqn:wealth-update} indicates that
the wealth must decay in order to forget the old earnings from
rejections far in the past. If we were to keep the first term and drop
the second, then the effect of the initial wealth (not just the
post-rejection earnings) also decays to zero. Intuitively, the
correction from the second term suggests that even if one forgets all
the past post-rejection earnings, the algorithm should behave as if it
started from scratch, which means that its initial wealth should not
decay. This does not contradict the fact that initial wealth can be
\emph{consumed} because of testing penalties $\phi_t$, but it should
not \emph{decay} with time---the decay was only introduced to avoid
piggybacking, which is an effect of post-rejection \emph{earnings} and
not the \emph{initial} wealth.

A natural
restriction on $\phi_t$ is the bound
$
\phi_t \leq \decay W(t-1) + (1-\decay)W_0 \One{\tau_1 > t - 1},
$
which ensures that the wealth stays non-negative.  Further,
$w_t,u_t,\alpha_t,\phi_t \in \F^{t-1}$ must be chosen so that the
rejection reward $\psi_t$ obeys conditions~\eqref{eqn:weighted-reward}
and~\eqref{eqn:weighted-reward2}.  Notice that setting $w_t = u_t =
\delta = 1$ recovers the GAI++ updates. 
As an example, mem-LORD++ would use :
\begin{align*}
\alpha_t = \gamma_{t} W_0 \decay^{t - \min\{\tau_1, t\}} + \sum_{j: \tau_j < t}  \decay^{t-\tau_j} \gamma_{t-\tau_j} \psi_{\tau_j}.
\end{align*}
We are now ready to present our last main result. 

\vspace{0.05in}
\begin{theorem}
  \label{thm:gai++memfdr}
Under independence, the doubly-weighted mem-GAI++ algorithm satisfies the bound 
$\EE{\dotfrac{\Vd_u(T) + W(T)}{\Rd_u(T)}} \leq \alpha \textnormal{ for
  all $T \in \mathbb{N}$}$. Since $W(T) \geq 0$, we have $\memfdr_u(T) \leq
\alpha$ for all $T \in \mathbb{N}$.
\end{theorem}

See~\appref{thmproof} for the proof of this
claim. \appref{alpha-death} discusses how to use ``abstaining'' to provide a smooth restart from alpha-death, whereas
\appref{sims} contains a numerical simulation demonstrating the use of
decaying memory.


\subsection{Abstinence for recovery from alpha-death}\label{app:alpha-death}

For truly temporal applications as outlined in motivation M2, we allow
the algorithm to \emph{abstain} from testing, meaning that it does not
need to perform a test at each time step.  In this case, we use the
convention of $P_t = -1$ to indicate that we abstained from testing at
time $t$. Also, we introduce the random variables
\begin{equation}
A_t \defn \One{P_t = -1}, ~~\text{ and }~~ A_t^c \defn 1 - A_t,
\end{equation}
as indicators for abstention. Abstention may happen due to the
natural variation in frequency of testing hypotheses in real-world
applications.
Additionally, abstention is the natural treatment for recovery from
alpha-death.  If the alpha-wealth is deemed too low, abstaining for a
while can drop $\memfdr$ below a threshold, and when it becomes small
enough, one can reset all variables and restart the entire process.
In more detail, note that we would change the quantities
$V(t),W(t),R(t)$ only if we actually did not abstain and performed a
test, as given by:
\begin{align*}
W(t) & \defn \decay W(t-1) + (1-\decay)W_0\One{\tau_1 > t - 1} - A_t^c
\phi_t + A_t^c R_t \psi_t \label{eqn:wealth-update-abst} \\
\Vd_u(t) & \defn \decay \Vd_u(t-1) + u_t A_t^c R_t \One{t \in \nulls}
\\
\Rd_u(t) &\defn \decay \Rd_u(t-1) + u_t A_t^c R_t .
\end{align*}
When we abstain, assuming that we have made at least one rejection,
all three quantities decay with time. Hence, the ratio
$\dotfrac{\Vd_u(t) + W(t)}{\Rd_u(t)}$ remains unchanged initially, 
and when the denominator $\Rd_u(t)$ falls below one,
the aforementioned ratio decays smoothly to zero (and hence so does
the $\memfdr_u$). Using a user-defined tolerance $\epsilon$,
we can then ``reset'' when $\Rd_u(t) < \epsilon$ by re-defining all
quantities to their starting values, setting the time to zero,
and restarting the entire process.

An alternative to abstinence is to pre-define a period of time after
which the process will reset, like a calendar year, or a single
financial quarter.  With this choice, decaying memory may help with
piggybacking but is not needed for recovery from alpha-death. However,
for applications in which there is no natural special period, and
which is in some sense continuous in time without discrete
breakpoints, the decaying memory FDR is a natural quantity to control,
and abstinence is an arguably intuitive solution to
alpha-death. Indeed, companies are obviously less willing to accept a
permanent alpha-death that ends all testing forever, and are more
likely to be willing to abstain from testing for a while, and run an
internal check on why they lost alpha-wealth by testing too many
nulls, or perhaps why they had very low signal on their non-nulls
(making them seem like nulls).


\section{Numerical Simulations}\label{app:sims}

Here, we provide proof-of-concept experiments for
various aspects of the paper.\footnote{The code for reproducing all experiments in this paper is publicly available at \texttt{https://github.com/fanny-yang/OnlineFDRCode}.}

\subsection{Evidence of higher power of GAI++ over GAI}

To demonstrate an improvement of GAI++ over GAI, we follow the simple
experimental setup of~\citet{javanmard2016online} which tests the
means of $T = 1000$ Gaussian distributions. The null hypothesis is
$H_j : \mu_j = 0$ for $j = \{1,\dots,T\}$. We observe independent $Z_j
\sim N(\mu_j, 1)$, which can be converted using the Gaussian CDF
$\Phi$ to a one-sided $p$-value, $P_j = \Phi(-Z_j)$ or to a two-sided
$p$-value, $P_j = 2\Phi(-|Z_j|)$. Notice that the $p$-value is exactly
uniformly distributed when the null hypothesis is true, that is
$\mu_j=0$. The means $\mu_j$ are set independently according to the
mixture:
\[
\mu_j \sim \begin{cases} 0 \text{ with probability } 1-\pi_1,
  \\ N(0,\sigma^2) \text{ with probability } \pi_1, \end{cases}
\]
and we set $\sigma^2 = 2\log T$, resulting in means that are near the
boundary of detectability. Indeed, under the global null where
$\pi_1=0$, $\max_j Z_j = (1+o(1))\sqrt{2 \log T}$, and $\sqrt{2 \log
  T}$ is the minimax amplitude for estimation under the sparse
Gaussian sequence model.

The improvement in power of GAI++ over GAI depends on the choice of
$W_0$ and $B_0 = \alpha - W_0$. If $W_0$ is too small, the algorithm may
suffer from alpha-death too quickly, because the signals may not be
strong enough for the algorithm to recover by accumulating the large
rewards $B_0$. If $W_0$ is too large, the reward $B_0$ at each step
will be too small, and the algorithm will suffer from lower
power. Hence, the larger $W_0$ is, the smaller $B_0$ is, and the more
GAI++ will improve over GAI. For our simulations, we set $W_0 =
\alpha/5$, for which we only expect a small improvement, and always
have $\alpha = 0.05$, and run 200 independent trials to estimate the
power and FDR.

\begin{figure}[h!]
\centering
   \begin{tabular}{ccc}
  \includegraphics[width=0.4\textwidth,trim={0 0 0
      0},clip]{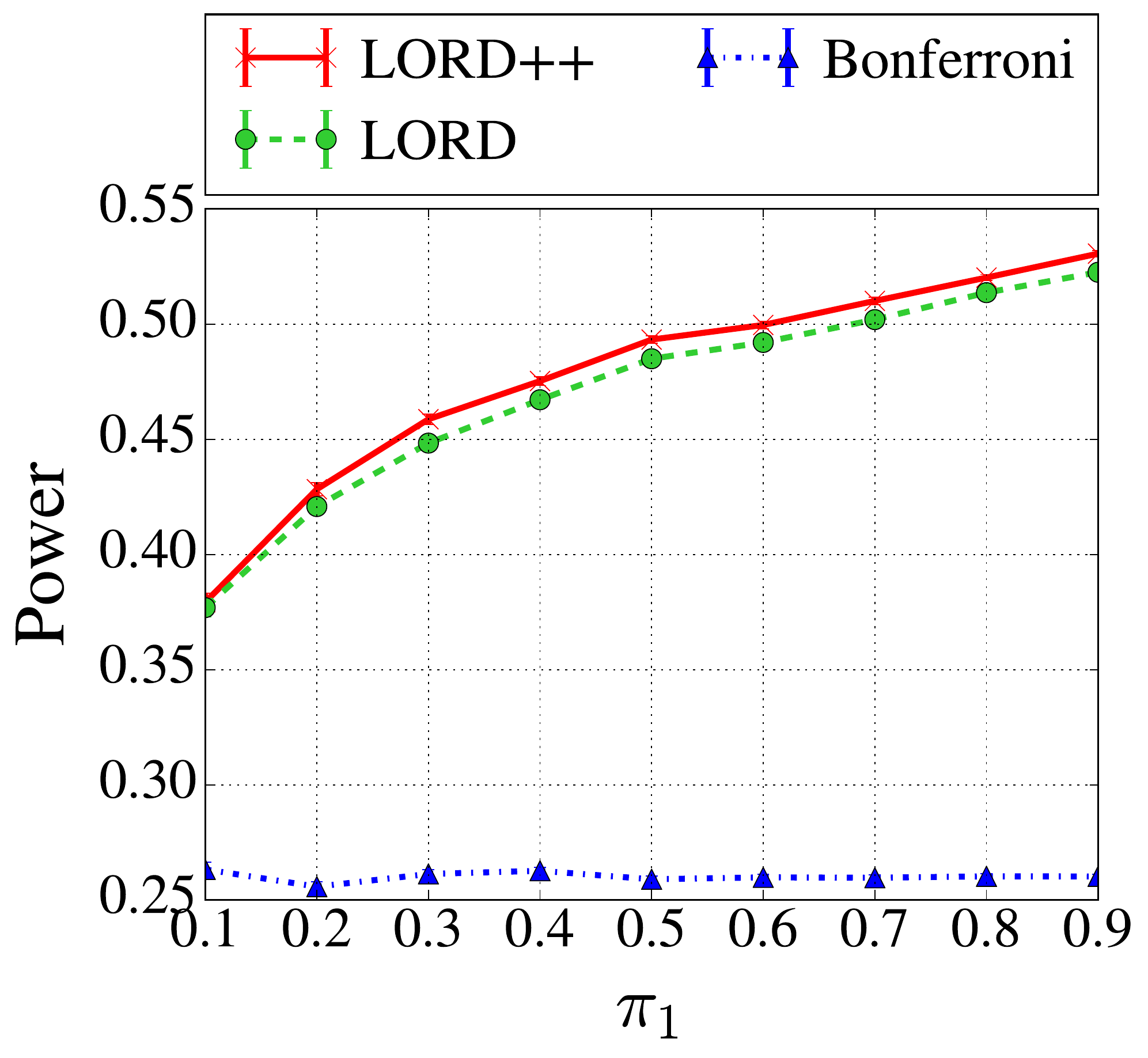} & \hspace{.5cm} &
  \includegraphics[width=0.4\textwidth]{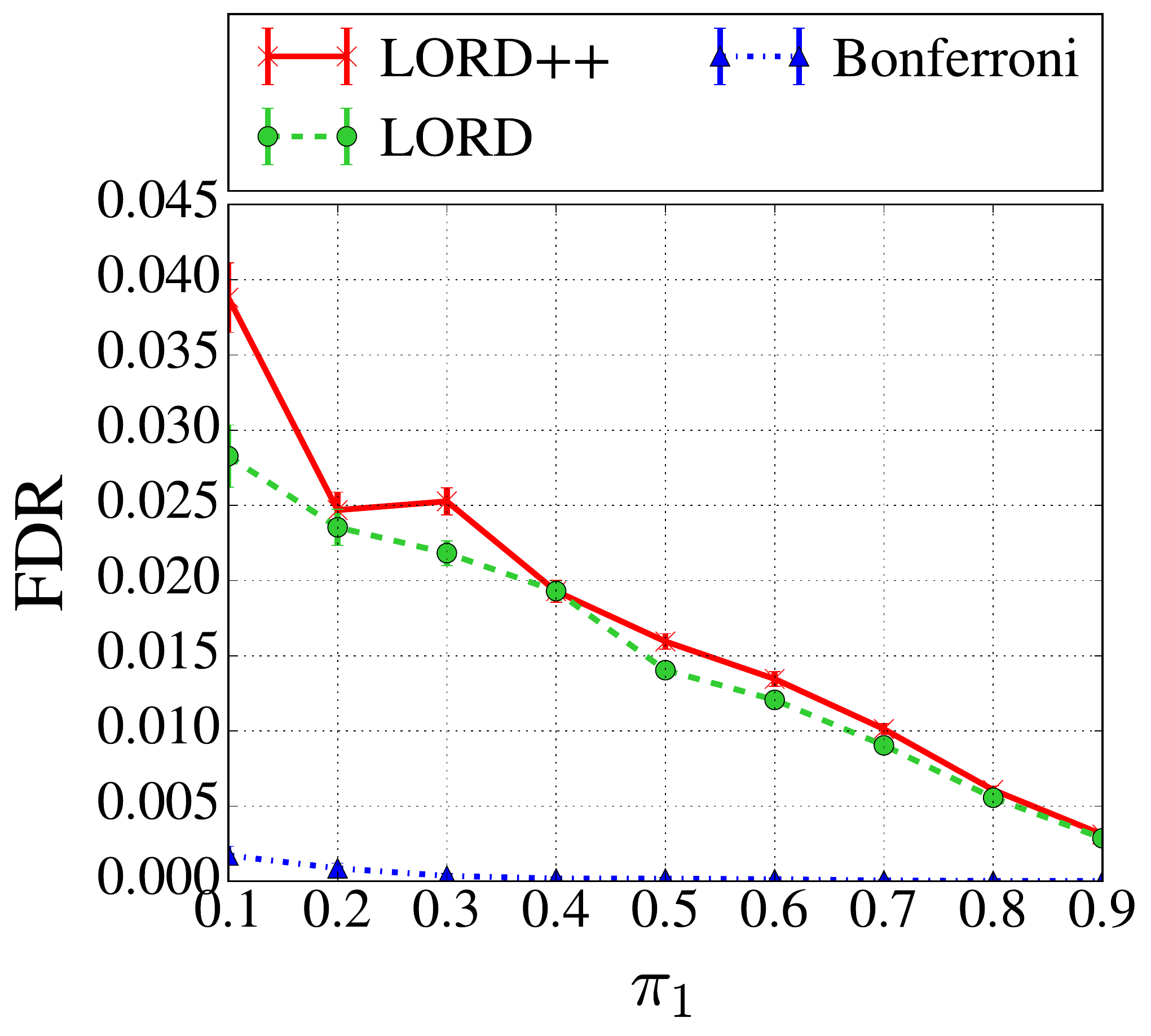} 
%
   \end{tabular}
  \caption{Plots of power vs $\pi_1$ (left panel) and FDR versus
    $\pi_1$ (right panel, for various algorithms.}
  \label{fig:GAI++}
\end{figure}

For concrete monotone GAI and GAI++ procedures, we choose LORD'17 as detailed
in Table~\ref{tab:examples} and LORD++ from definition \eqref{eqn:LORD++}.
We define power as usual in the FDR literature:
\begin{equation*}
 \power(T) \defn  \EE{\dotfrac{\sum_{t\in \nonnulls} R_t}{\sum_{t=1}^T \One{t \in \nonnulls}}}.
\end{equation*}

In \figref{GAI++}, we plot the power and FDR for the Bonferroni and
LORD algorithms using $W_0 = \alpha/2$ and the constant sequenuce $\gamma_j = 0.0722 \frac{\log( j \vee 2)}{j e^{\sqrt {\log j}}}$
derived for testing Gaussian means \cite{javanmard2016online},
where the leading constant was approximated so that the infinite sequence sums
to one.  As predicted by the theory, the power of the LORD++ algorithm is uniformly
better than LORD.


\subsection{Piggybacking and decaying memory}\label{app:piggyplots}

For this subsection, we move away from the stationary setting that is
a useful base case, but unrealistic in practice. To bring out the
phenomenon of piggybacking, we consider the setting where $\pi_1 \gg
0.5$ in the first 1000 tests, and $\pi_1 \ll 0.5$ in the second
1000. There is nothing specific to this particular choice, and will
qualitatively occur whenever there is a stretch of non-nulls followed by a
stretch of nulls. For simplicity, we restrict our attention to the
LORD++ and the mem-LORD++ algorithms, and plot their $\memfdr$ as a
function of time.  In particular, we use the following concrete update 
for $\alpha_t$ in the mem-LORD++ algorithm:
\begin{align*}
\alpha_t = 
\gamma_t W_0 \decay^{t - \min\{\tau_1, t\}} 
+ (\alpha - W_0) \decay^{t-\tau_1} \gamma_{t-\tau_1} 
+ \alpha \: \big(\sum_{\tau_j < t, \tau_j \neq \tau_1}
\decay^{t-\tau_j} \gamma_{t-\tau_j} \big).
\end{align*}

\begin{figure}[h!]
  \begin{center}
    \begin{tabular}{ccc}
   \includegraphics[width=0.4\textwidth,trim={0 0 0 0},clip]{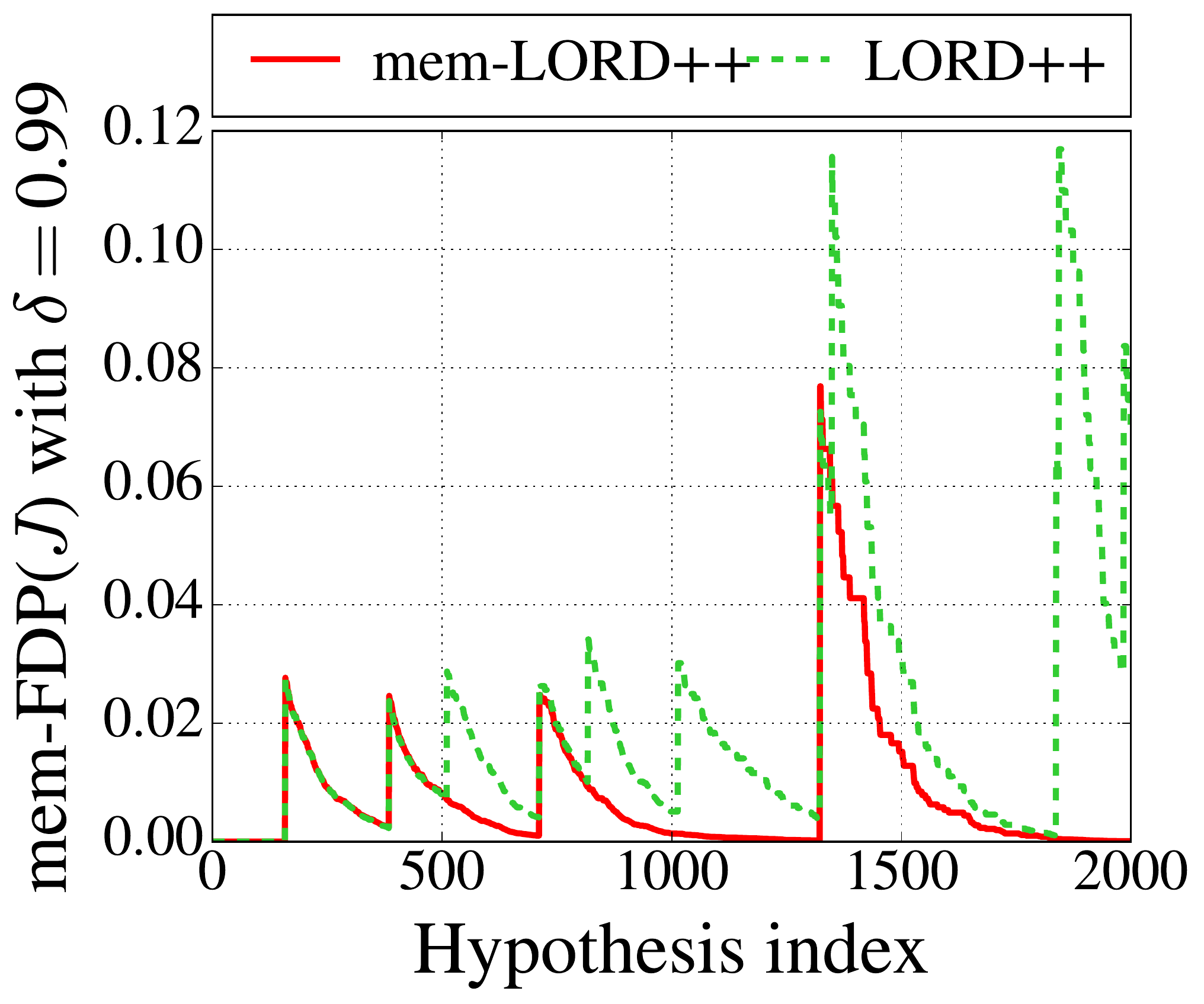} & \hspace{.5cm} &
  \includegraphics[width=0.4\textwidth]{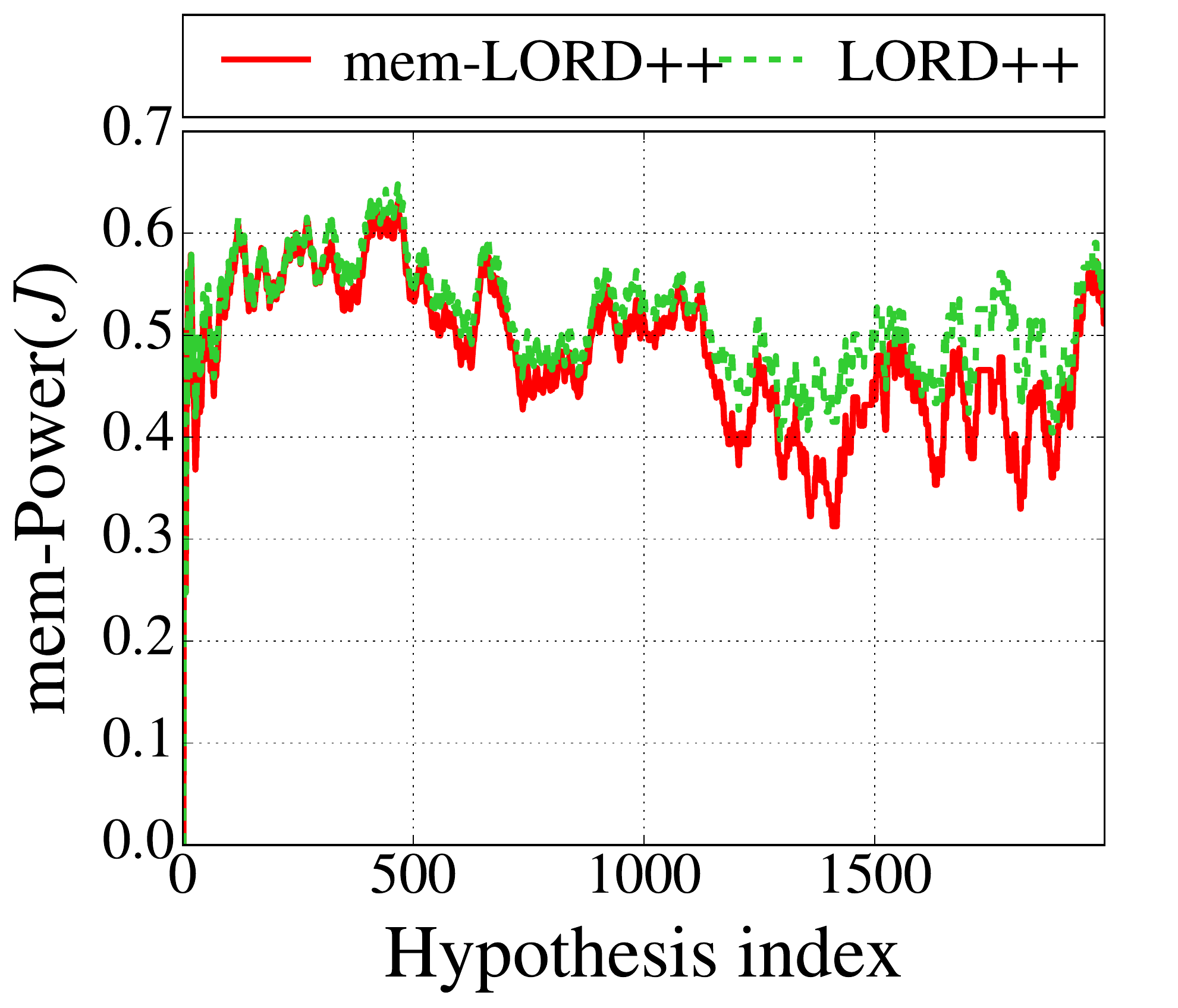}  
    \end{tabular}
  \end{center}
  \caption{Plots of mem-power versus time (left panel and mem-FDR
    versus time (right panel), for LORD++ and mem-LORD++ with
    $\delta=0.99$. The spike in false discoveries suffered by LORD++
    due to piggybacking is significantly smoothed by mem-LORD++
    without much loss of power.}
  \label{fig:piggy}
\end{figure}

\figref{piggy} demonstrates see that LORD++ suffers a large spike in $
\memfdr$ locally in time, which is significantly smoothed out by
mem-LORD++ with $\delta = 0.99$, at an insignificant loss of
power. Arguably, the power should itself be replaced by a ``decaying
memory power'' which we call mem-power, which definition is analogous
to $\memfdr$ in relation to FDR, i.e. 
\begin{equation*}
\small
\mempower(T) \defn \EE{\dotfrac{\tVd(T)}{\tRd(T)}},
\end{equation*}
where $\tVd(T) \defn \delta \tVd(T-1) + R_T\One{T\in \nonnulls}$ and
$\tRd(T) \defn \delta \tRd(T-1) + \One{T\in \nonnulls}$.
Due to its conservative choice for $\alpha_t$, the smoothing of the
$\memfdr$ measure comes at the expense of lower mem-power for mem-LORD++
compared to LORD++ in the second half of the experiment.


\subsection{Alpha-death}
\label{app:adplots}
\begin{figure}[h!]
  \begin{center}
    \begin{tabular}{ccc}
   \includegraphics[width=0.4\textwidth,trim={0 0 0 0},clip]{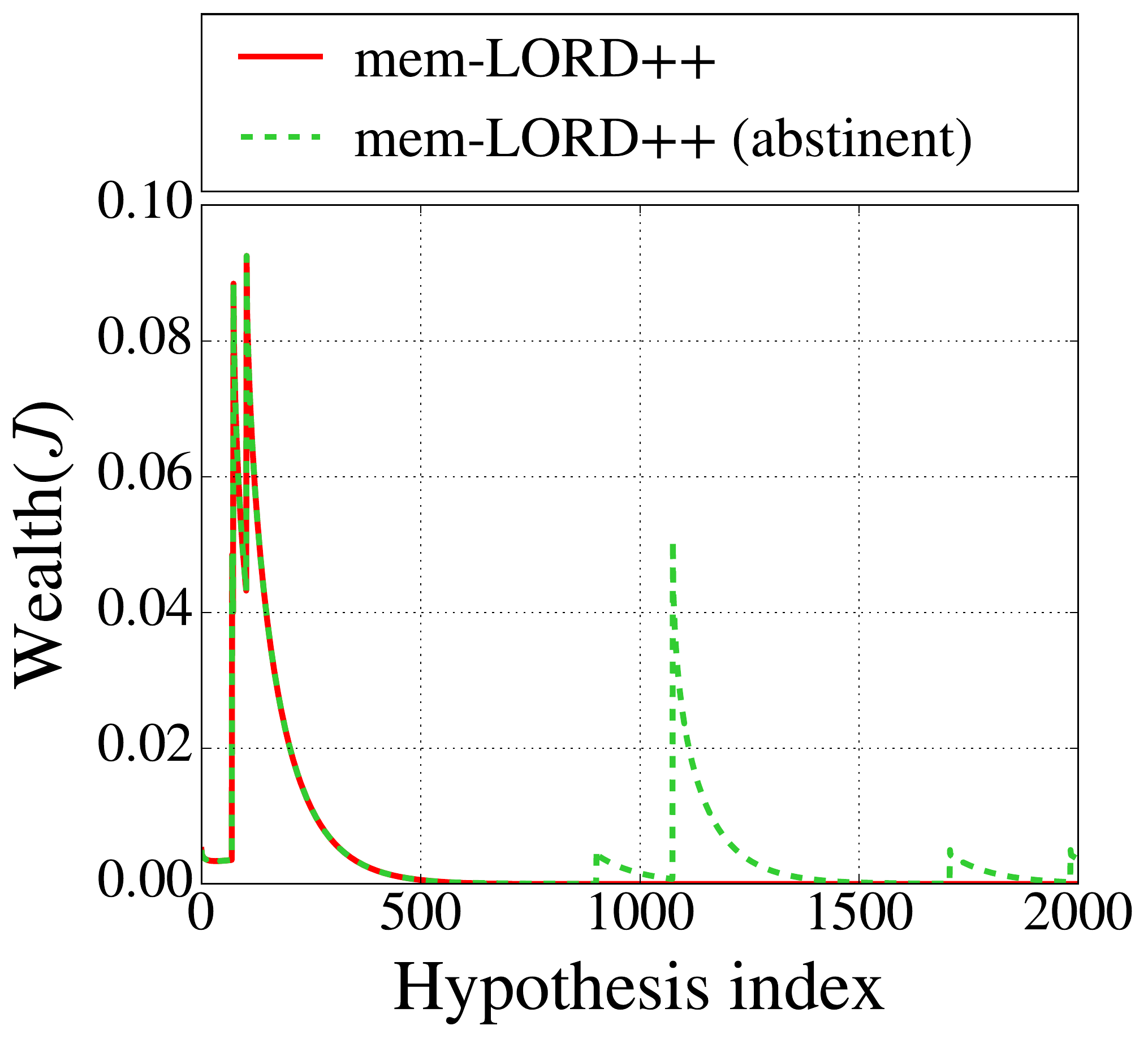} & \hspace{.5cm} &
  \includegraphics[width=0.4\textwidth]{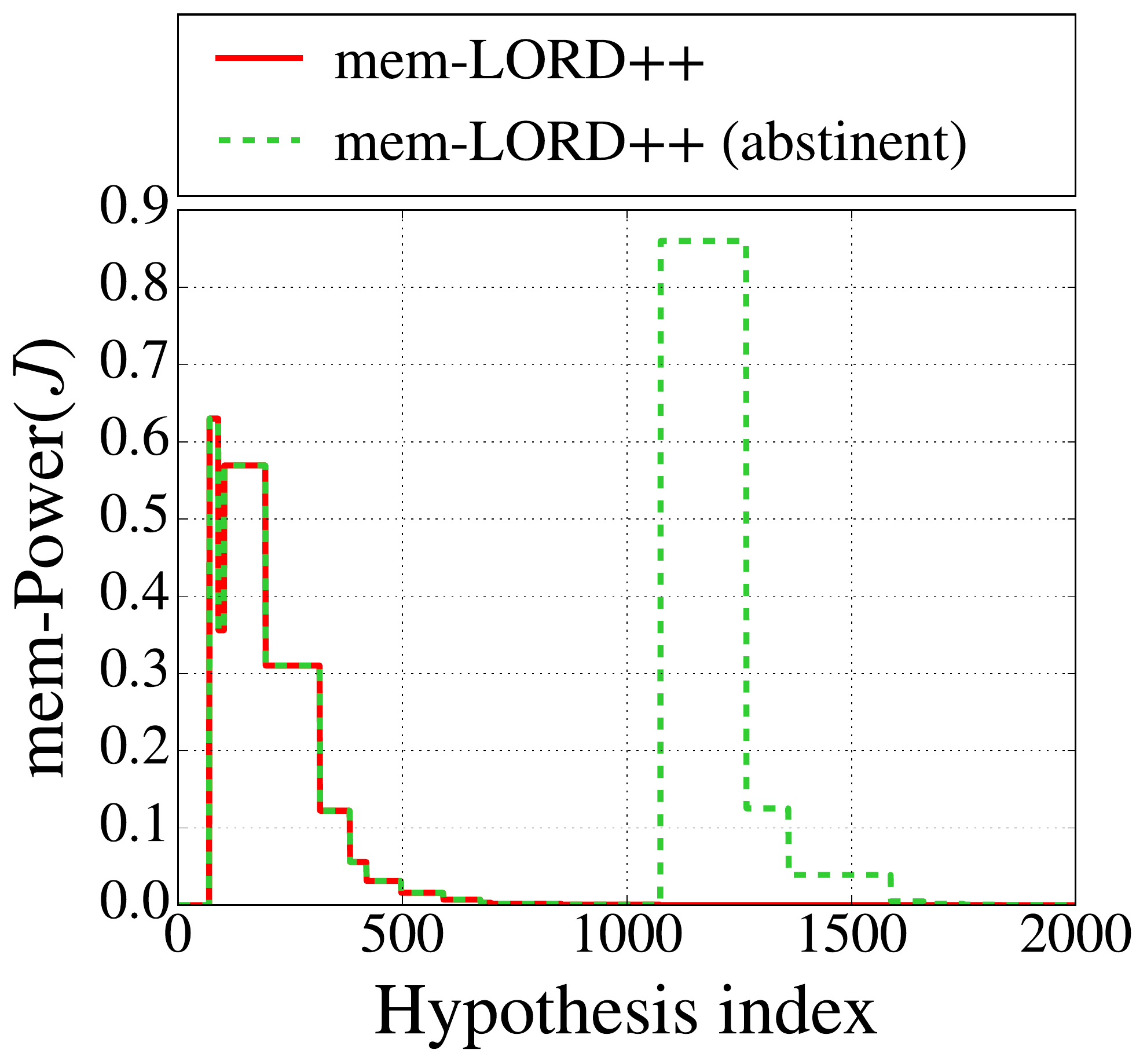}  
    \end{tabular}
  \end{center}
  \caption{Plots of wealth versus time (left panel)  power
    versus time (right panel), for mem-LORD++ with
    $\delta=0.99$ and a constant $\pi_1 = 0.01$. Once the wealth vanishes, 
    the generic mem-LORD++ cannot make new discoveries for the entire future, whereas the
    abstinent mem-LORD++ circumvents this issue and eventually starts anew, allowing
    new incoming non-nulls to be detected.}
  \label{fig:alphadeath}
\end{figure}

Here, we illustrate the usefulness of abstinence
as discussed in Section~\ref{app:alpha-death} for experiments
where alpha-death is reached rather quickly. Concretely, we choose the
probability of each hypothesis being non-null to be identically and
independently $p = 0.01$. Furthermore, we abstain 
from testing if $W(t) < \eps_w$ and we reset to initial values if
$R(t) < \eps_r$ with $\eps_w = 0.05 W_0$ and $\eps_r = 0.1$. 
\figref{alphadeath} depicts both the time development of wealth on the
left hand side and the corresponding mem-power on the right hand side.
The red curves representing the generic mem-LORD++ algorithm show that
once wealth reaches $0$, no discoveries can be made so that mem-power
stays at $0$ for the entire rest of the experiment. On the other hand,
for the exact same experiment, the abstinent mem-LORD++ in green has a ``second
chance'' after abstaining for a while: the experiment is reset
so that new discoveries can be made even though the wealth had depleted
at some previous time.

\subsection{Subtleties with the use of prior weights}
\label{app:weights}

If one has a high prior belief that a hypothesis is non-null, then 
the ``oracle'' strategy of assigning weights depends on the strength
of the underlying signal: (a) if the signal is small, an oracle would
assign a weight that is just high enough to reject the non-null, while
earning a small reward, and (b) if the signal is large, then an oracle
would assign a weight as small as possible to just reject the
non-null, earning as large a reward as possible, amassing alpha-wealth
to be used for later tests.

\begin{figure}[h!]
\centering
  \begin{tabular}{cc c}
  \includegraphics[width=0.4\textwidth,trim={0 0 0
      0},clip]{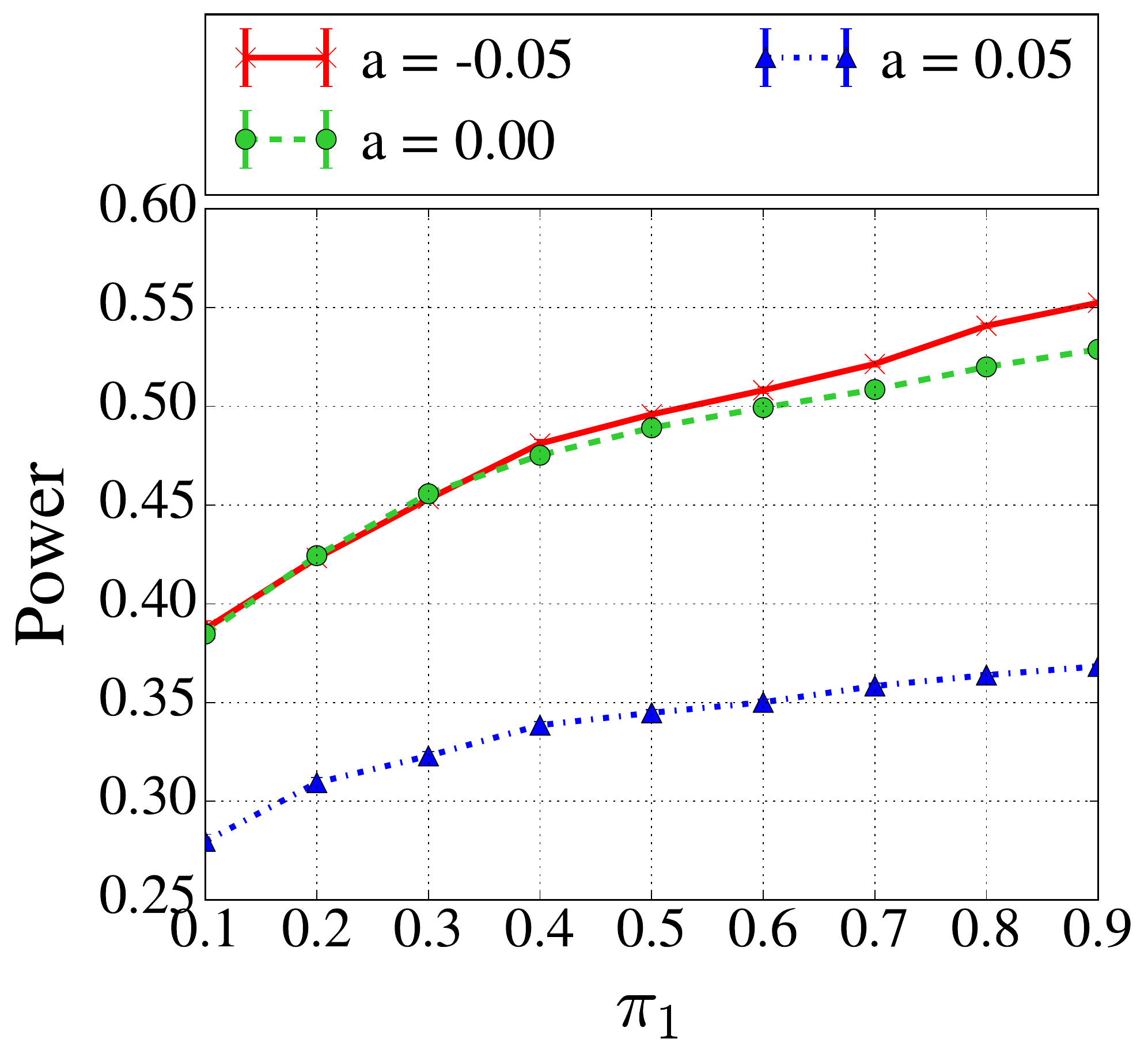} & \hspace{.5cm}&
  \includegraphics[width=0.4\textwidth]{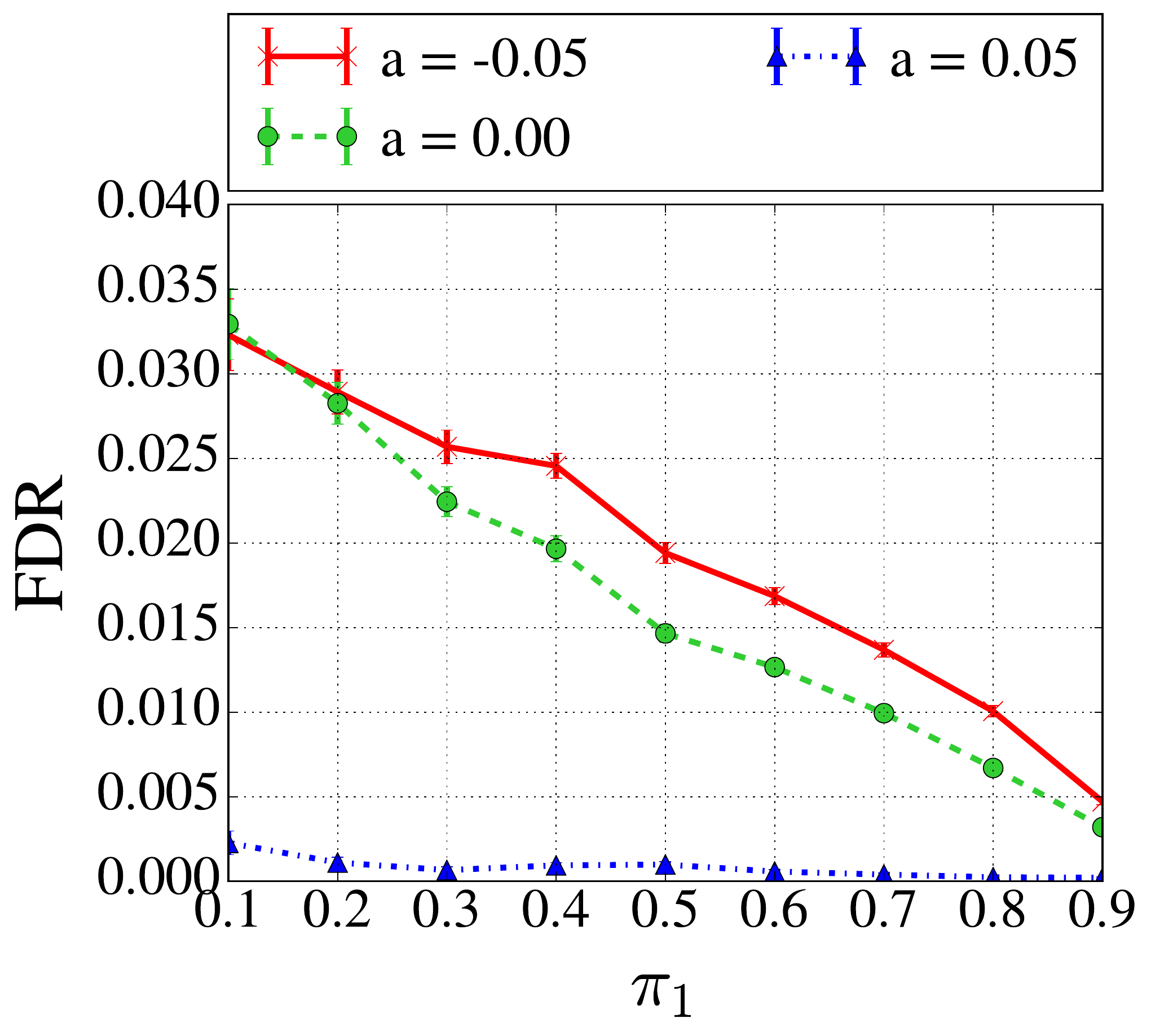} 
   \end{tabular}
  \caption{Plots of power vs $\pi_1$ (left panel) and FDR versus
    $\pi_1$ (right panel), for LORD++ with weights $1+a$ on non-nulls
    and $1-a$ on nulls.}
  \label{fig:weights}
\end{figure}

\figref{weights} suggests that in the aforementioned simulation setup,
we happened to be in situation (b), where most non-nulls had enough
signal so that using a weight smaller than one was more beneficial than a
weight larger than one. We used the same setup as the previous
subsection, except that we assign ``oracle'' weights of $1+a$ whenever the
hypothesis is non-null, and a weight of $1-a$ whenever the hypothesis
is null, for positive and negative choices of $a$. We
 use the word ``oracle'' since, in practice, we of course do not know which
  hypotheses are null and non-null. 

\section{Proofs}

In this section, we organize the various proofs of lemmas and theorems found in the paper.

\subsection{Proof of~\lemref{power}}
\label{app:lemproof}

Letting $\vec{P} = (P_1,\ldots,P_T)$ be the original vector of
$p$-values, we define a ``hallucinated'' vector of $p$-values
$\widetilde{P}^{-t} \defn (\widetilde P_1,\ldots,\widetilde P_T)$ that equals $\vec{P}$, except that
the $t$-th component is set to zero : 
\begin{align*}
  \widetilde P_i = \begin{cases} 0 & \mbox{if $i = t$} \\ P_i &
    \mbox{if $i \neq t$.}
  \end{cases}
\end{align*}
For all $i$, define $\widetilde R_i = \One{\widetilde P_i \leq
  f_i(\widetilde R_1,\ldots,\widetilde R_{i-1})}$ and let the
corresponding vectors of rejections using $\vec{P}$ and $\widetilde{P}^{-t} $ be $\vec{R} = (R_1,\ldots,R_T)$
and $\widetilde{R}^{-t} = (\widetilde R_1,\ldots, \widetilde R_T)$. By
construction, we have $\widetilde{R}_i = R_i$ for all $i < t$, and
$\widetilde{R}_i \geq R_i$ for all $i \geq t$, from which we conclude
that $f_i(R_1,\ldots,R_{i-1}) = f_i(\widetilde R_1,\ldots,\widetilde
R_{i-1})$ for all $i \leq t$.
Also,
we know $\widetilde R_t = 1$ by construction since $\widetilde P_t =
0$ implying that $g(\widetilde{R}^{-t}) > 0$. Hence, on the event
$\{P_t \leq f_t(R_1,\ldots,R_{t-1})\}$, we have $R_t = \widetilde R_t
= 1$ and hence also $\vec{R} = \widetilde{R}^{-t}$. This allows us to conclude
that
\[
\dotfrac{\One{P_t \leq f_t(R_1,\ldots,R_{t-1})}}{g (\vec{R})} =
\dotfrac{\One{P_t \leq f_t(R_1,\ldots,R_{t-1})}}{g
  (\widetilde{R}^{-t})}.
\]

Since $\widetilde{R}^{-t}$ is independent of $P_t$, we may take
conditional expectations to obtain
\begin{align*}
\EEst{\dotfrac{\One{P_t \leq f_t(R_1,\ldots,R_{t-1})}}{g (\vec{R})}
}{\F^{t-1}} &= \EEst{ \dotfrac{\One{P_t \leq
      f_t(R_1,\ldots,R_{t-1})}}{g (\widetilde{R}^{-t})} }{ \F^{t-1}}
\\ &\stackrel{(i)}{\leq} \EEst{ \dotfrac{ f_t(R_1,\ldots,R_{t-1})}{g
    (\widetilde{R}^{-t})} }{ \F^{t-1}} \\ &\stackrel{(ii)}{\leq}
\EEst{ \dotfrac{ f_t(R_1,\ldots,R_{t-1})}{g (\vec{R})} }{ \F^{t-1}},
\end{align*}
where inequality (i) follows by taking expectation only with respect
to $P_t$ by invoking the conditional super-uniformity
property~\eqref{eqn:superuniformity-cond}; and inequality (ii) follows
because $g (\vec{R}) \leq g(\widetilde{R}^{-t})$ since $R_i \leq
\widetilde R_i$ for all $i$ by monotonicity of the online FDR rule.  
This concludes the proof of the lemma.


\subsection{Proof of \thmref{newview}}\label{app:proof-newview}

For any time $T \in \N$, we may infer that mFDR is controlled using the following argument :
\begin{align*}
\EE{V(T)} &= \sum_{j \in \nulls, j \leq T} \EE{\EEst{\One{P_j \leq \alpha_j}}{\F^{j-1}}} \\
&\stackrel{(i)}{\leq} \sum_{j \in \nulls, j \leq T} \EE{\alpha_j} \\
&\stackrel{(ii)}{\leq} \EE{\sum_{j \leq T} \alpha_j} \\
&\stackrel{(iii)}{\leq} \alpha \EE{R(T) \vee 1}, 
\end{align*}
where inequality $(i)$ follows after taking iterated expectations by conditioning on $\F^{j-1}$, and then applying the conditional superuniformity property \eqref{eqn:superuniformity-cond}, inequality $(ii)$ follows simply by dropping the condition $j \in \nulls$, and inequality $(iii)$ follows by the theorem assumption  that $\frac{\sum_{j\leq T}\alpha_j}{R(T)} \leq \alpha$. Rearranging yields the conclusion $\mfdr(T) = \dotfrac{\EE{V(T)}}{\EE{R(T)}} \leq \alpha$, as desired.

When the sequence $\{\alpha_t\}$ is additionally monotone, we can use the following argument to prove that the procedure controls FDR at any time $T \in \N$ :
\begin{align*}
\fdr = \EE{\dotfrac{V(T)}{R(T)}} &= \sum_{j \in \nulls, j \leq T} \EE{\dotfrac{\One{P_j \leq \alpha_j}}{R(T)}} \\
&\stackrel{(iv)}{\leq} \sum_{j \in \nulls, j \leq T} \EE{\dotfrac{\alpha_j}{R(T)}} \\
&\stackrel{(v)}{\leq} \EE{\dotfrac{\sum_{j \leq T} \alpha_j}{R(T)}} \\
&\stackrel{(vi)}{\leq} \alpha,
\end{align*}
where inequality $(iv)$ follows after taking iterated expectations by conditioning on $\F^{j-1}$, and then applying the conditional superuniformity lemma \lemref{power}, and inequalities $(v)$ and $(vi)$ follow for the same reasons as inequalities $(ii)$ and $(iii)$. 

This concludes the proof of both parts of the theorem.


\subsection{Proof of~\thmref{gai++}}\label{app:thmproof++}


Substituting the definitions of $V(T) = \sum_{t=1}^T R_t \One{t \in
  \nulls}$ and the alpha-wealth
\[
W(T) = W_0 + \sum_{t=1}^T  (-\phi_t + R_t \psi_t),
\]
we may use the tower property of conditional expectation to write 
\begin{align*}
\EE{\dotfrac{V(T) + W(T)}{R(T)}} = &\sum_t \underbrace{\EE{\EEst{ \dotfrac{\frac{W_0}{T} + R_t(\psi_t + \One{t \in \nulls}) - \phi_t}{R(T)} }{\F^{t-1} }} }_{L_t}.
\end{align*}
We tackle the above expression term by term, depending on whether or not $t \in \nulls$.

\paragraph{Case 1.}  First, suppose that $t \in \nulls$.
Substituting $\psi_t \leq \frac{\phi_t }{ \alpha_t} + b_t - 1$ into
the expression for $L_t$ yields
\begin{align}
L_t & \leq \EE{\EEst{ \dotfrac{\frac{W_0}{T} + R_t(\frac{\phi_t
      }{\alpha_t} + b_t ) - \phi_t}{R(T)} }{\F^{t-1} }} \nonumber
\\
& = \EE{\EEst{ \dotfrac{\frac{W_0}{T} +R_t b_t +
      \frac{\phi_t}{\alpha_t} ( R_t - \alpha_t )}{R(T)} }{\F^{t-1}
}}, \label{eqn:midproof1}
\end{align}
where the equality follows simply by rearrangement.  Since $t \in
\nulls$, invoking~\lemref{power} guarantees that
\begin{equation}
  \label{eqn:bound1}
\EEst{ \dotfrac{R_t}{R(T)}}{\F^{t-1}} = \EEst{ \dotfrac{\One{P_t \leq
      \alpha_t }}{R(T)}}{\F^{t-1}} \leq \EEst{\dotfrac{\alpha_t
  }{R(T)} }{\F^{t-1}},
\end{equation}
since the two mappings $(R_1,\dots,R_T) \mapsto R(T)$ and
$(R_1,\dots,R_{t-1}) \mapsto \alpha_t \in \F^{t-1}$ are coordinatewise
nondecreasing, as required to apply \lemref{power}. Since
$\phi_t,\alpha_t$ are $\F^{t-1}$-measurable,
equation~\eqref{eqn:bound1} implies that the last term in the
numerator of equation~\eqref{eqn:midproof1} is negative, and hence
\begin{align*}
L_t \leq \EE{\EEst{ \dotfrac{\frac{W_0}{T} + R_t b_t }{R(T)} }{\F^{t-1} }}.
\end{align*}

\paragraph{Case 2.}  Now suppose that  $t \notin \nulls$. Substituting
$\psi_t \leq \phi_t + b_t$ into the expression for $L_t$ yields
\begin{align}
L_t & \leq \EE{\EEst{ \dotfrac{\frac{W_0}{T} + R_t(\phi_t + b_t) -
      \phi_t}{R(T)} }{\F^{t-1} }} \nonumber \\ &= \EE{\EEst{
    \dotfrac{\frac{W_0}{T} + R_t b_t + \phi_t (R_t - 1)}{R(T)}
  }{\F^{t-1} }}, \nonumber
\end{align}
where the equality follows simply by rearrangement. Since $R_t \leq
1$, we may infer that
\begin{align*}
L_t \leq \EE{\EEst{ \dotfrac{\frac{W_0}{T} + R_t b_t}{R(T)} }{\F^{t-1} }},
\end{align*}
which is the same expression as the bound derived in Case 1.

\paragraph{Combining both cases.} We complete the proof by combining
the two cases. Using the same bound for $L_t$ in both cases yields
\begin{align}
  \label{eqn:midproof2}
\EE{\dotfrac{V(T) + W(T)}{R(T)}} &\leq \EE{\dotfrac{W_0 + \sum_t R_t
    b_t}{R(T)} }.
\end{align}
We now note that $b_t$ always equals $\alpha$, except for the very
first rejection at time $\tau_1$, in which case it equals $\alpha -
W_0$. Hence, we may have $\sum_t R_t b_t = \sum_t R_t \alpha - W_0
\One{T \geq \tau_1}$.  Substituting this expression into the
bound~\eqref{eqn:midproof2} yields
\begin{align*}
\EE{\dotfrac{V(T) + W(T)}{R(T)}} &\leq \EE{\dotfrac{W_0 + \alpha R(T)
    - W_0 \One{T \geq \tau_1} }{R(T)} } \leq \alpha,
\end{align*}
which completes the proof of the theorem.


\subsection{Proof of~\thmref{gai++weighted}}\label{app:thmproofweights}

Substituting the definitions of $V_u(T) = \sum_{t=1}^T u_t R_t \One{t
  \in \nulls}$ and the alpha-wealth
\begin{align*}
W(T) = W_0 + \sum_{t=1}^T (-\phi_t + R_t \psi_t),
\end{align*}
we may use the tower property of conditional expectation to write 
\begin{align*}
\EE{\dotfrac{V_u(T) + W(T)}{R_u(T)}} = &\sum_t \underbrace{\EE{\EEst{
      \dotfrac{\frac{W_0}{T} + R_t(\psi_t + u_t\One{t \in \nulls}) -
        \phi_t}{R_u(T)} }{\F^{t-1} }} }_{L_t}.
\end{align*}
We tackle the above expression term by term, depending on whether or
not $t \in \nulls$.

\paragraph{Case 1.}  First suppose that  $t \in \nulls$.
Substituting $\psi_t \leq \frac{\phi_t}{u_t w_t \alpha_t} +
u_t b_t - u_t$ into the expression for $L_t$ yields
\begin{align}
L_t & \leq \EE{\EEst{ \dotfrac{\frac{W_0}{T} + R_t(\frac{\phi_t
        }{u_t w_t \alpha_t} + u_t b_t ) - \phi_t}{R_u(T)}
  }{\F^{t-1} }} \nonumber \\ &= \EE{\EEst{ \dotfrac{\frac{W_0}{T} +R_t
      u_t b_t +\frac{\phi_t }{u_t w_t \alpha_t} ( R_t -
      \alpha_t w_t u_t)}{R_u(T)} }{\F^{t-1}
}}, \label{eqn:weightedmidproof1}
\end{align}
where the equality follows simply by rearrangement.  Since $t \in
\nulls$, by invoking~\lemref{power}, we may infer that
\begin{align}
  \label{eqn:weightedbound1}
\EEst{ \dotfrac{R_t}{R_u(T)}}{\F^{t-1}} = \EEst{ \dotfrac{\One{P_t
      \leq \alpha_t w_t u_t}}{R_u(T)}}{\F^{t-1}}
\leq \EEst{\dotfrac{\alpha_t w_t u_t}{R_u(T)}
}{\F^{t-1}},
\end{align}
since the four mappings $(R_1,\dots,R_T) \mapsto R_u(T)$ and 
$(R_1,\dots,R_{t-1}) \mapsto \alpha_t,w_t,u_t $ are all coordinatewise
nondecreasing, as required to apply \lemref{power}. Since
$\phi_t,\alpha_t,w_t,u_t$ are $\F^{t-1}$-measurable,
equation~\eqref{eqn:weightedbound1} implies that the last term in the
numerator of equation \eqref{eqn:weightedmidproof1} is negative, and
hence
\begin{align*}
L_t \leq \EE{\EEst{ \dotfrac{\frac{W_0}{T} + R_tu_t b_t }{R_u(T)} }{\F^{t-1} }}.
\end{align*}


\paragraph{Case 2.}  Now suppose that $t \notin \nulls$.
Substituting $\psi_t \leq \phi_t + u_t b_t$ into the expression for
$L_t$ yields
\begin{align}
L_t & \leq \EE{\EEst{ \dotfrac{\frac{W_0}{T} + R_t(\phi_t + u_t b_t) -
      \phi_t}{R_u(T)} }{\F^{t-1} }} \nonumber \\ &= \EE{\EEst{
    \dotfrac{\frac{W_0}{T} + R_t u_t b_t + \phi_t (R_t - 1)}{R_u(T)}
  }{\F^{t-1} }}, \nonumber
\end{align}
where the equality follows simply by rearrangement. Since $R_t \leq
1$, we may infer that
\[
L_t \leq \EE{\EEst{ \dotfrac{\frac{W_0}{T} + R_t u_t b_t}{R_u(T)}
  }{\F^{t-1} }},
\]
which is the same expression as the bound derived in Case 1.

\paragraph{Combining both cases.} Finally, we combine the two cases.
Using the same bound for $L_t$ in both cases, and exchanging the
summation and expectation, we may conclude by definition of $b_t$ that
\begin{align}
  \label{eqn:weightedmidproof2}
\EE{\dotfrac{V_u(T) + W(T)}{R_u(T)}} &\leq \EE{\dotfrac{W_0 + \sum_t
    R_t u_t b_t}{R_u(T)} }.
\end{align}
We now note that $b_t$ always equals $\alpha$, except for the very
first rejection at time $\tau_1$, in which case it equals $\alpha -
\frac{W_0}{u_{\tau_1}}$. Hence, we may write $\sum_t R_t u_t b_t =
\sum_t R_t u_t \alpha - W_0 \One{T \geq \tau_1}$. Substituting the
above expression into the bound~\eqref{eqn:weightedmidproof2} yields
\begin{align*}
\EE{\dotfrac{V_u(T) + W(T)}{R_u(T)}} &\leq \EE{\dotfrac{W_0 + \alpha
    R_u(T) - W_0 \One{T \geq \tau_1} }{R_u(T)} } \leq \alpha,
\end{align*}
which completes the proof of the theorem.


\subsection{Proof of~\thmref{gai++memfdr}}
\label{app:thmproof}

Substituting the definitions of $\Vd_u(T) = \sum_{t=1}^T \decay^{T-t}
u_t R_t \One{t \in \nulls}$ and the alpha-wealth
\[
W(T) = W_0\decay^{T - \min\{\tau_1, T \}} + \sum_{t=1}^T \decay^{T-t}
(-\phi_t + R_t \psi_t),
\] 
we may use the tower property to write
\begin{align*}
\EE{\dotfrac{\Vd_u(T) + W(T)}{\Rd_u(T)}} = &\sum_t
\underbrace{\EE{\EEst{ \dotfrac{\frac{W_0}{T}\decay^{T - \min\{\tau_1,
          T \}} + \decay^{T-t}R_t(\psi_t + u_t\One{t \in \nulls}) -
        \decay^{T-t}\phi_t}{\Rd_u(T)} }{\F^{t-1} }} }_{L_t}.
\end{align*}
We tackle the above expression term by term, depending on whether or
not $t \in \nulls$.

\paragraph{Case 1} First suppose that $t \in \nulls$. Substituting
$\psi_t \leq \frac{\phi_t }{u_t w_t \alpha_t} + u_t b_t -
u_t$ into the expression for $L_t$ yields
\begin{align}
L_t & \leq \EE{\EEst{ \dotfrac{\frac{W_0}{T}\decay^{T - \min\{\tau_1,
        T \}} + \decay^{T-t}R_t(\frac{\phi_t }{u_t w_t
        \alpha_t} + u_t b_t ) - \decay^{T-t}\phi_t}{\Rd_u(T)}
  }{\F^{t-1} }} \nonumber \\ &= \EE{\EEst{
    \dotfrac{\frac{W_0}{T}\decay^{T - \min\{\tau_1, T \}} +
      \decay^{T-t}R_tu_t b_t + \decay^{T-t}
      \frac{\phi_t}{u_t w_t \alpha_t} ( R_t - \alpha_t
      w_t u_t )}{\Rd_u(T)} }{\F^{t-1}
}}, \label{eqn:decaymidproof1}
\end{align}
where the equality follows simply by rearrangement.  Since $t \in
\nulls$, by invoking \lemref{power}, we may infer that
\begin{align}
\EEst{ \dotfrac{R_t}{\Rd_u(T)}}{\F^{t-1}} = \EEst{ \dotfrac{\One{P_t
      \leq \alpha_t w_t u_t}}{\Rd_u(T)}}{\F^{t-1}}
\leq \EEst{\dotfrac{\alpha_t w_t u_t}{\Rd_u(T)}
}{\F^{t-1}},
\label{eqn:decaybound1}
\end{align}
since the four mappings $(R_1,\dots,R_T) \mapsto \Rd_u(T)$ and
$(R_1,\dots,R_{t-1}) \mapsto \alpha_t,w_t,u_t$ are coordinatewise
nondecreasing, as required to apply \lemref{power}. Since
$\phi_t,\alpha_t,w_t,u_t$ are $\F^{t-1}$-measurable,
equation~\eqref{eqn:decaybound1} implies that the last term in the
numerator of equation \eqref{eqn:decaymidproof1} is negative, and
hence
\begin{align*}
L_t \leq \EE{\EEst{ \dotfrac{\frac{W_0}{T}\decay^{T - \min\{\tau_1, T
        \}} + \decay^{T-t}R_tu_t b_t }{\Rd_u(T)} }{\F^{t-1} }}.
\end{align*}

\paragraph{Case 2} Next, suppose that $t \notin \nulls$. Substituting
$\psi_t \leq \phi_t + u_t b_t$ into the expression for $L_t$ yields
\begin{align}
L_t &\leq \EE{\EEst{ \dotfrac{\frac{W_0}{T}\decay^{T - \min\{\tau_1, T
        \}} + \decay^{T-t}R_t(\phi_t + u_t b_t) -
      \decay^{T-t}\phi_t}{\Rd_u(T)} }{\F^{t-1} }} \nonumber \\ &=
\EE{\EEst{ \dotfrac{\frac{W_0}{T}\decay^{T - \min\{\tau_1, T \}} +
      \decay^{T-t}R_t u_t b_t + \decay^{T-t}\phi_t (R_t -
      1)}{\Rd_u(T)} }{\F^{t-1} }}, \nonumber
\end{align}
where the equality follows simply by rearrangement. Since $R_t \leq
1$, we may infer that
\[
L_t \leq \EE{\EEst{ \dotfrac{\frac{W_0}{T}\decay^{T - \min\{\tau_1, T
        \}} + \decay^{T-t}R_t u_t b_t}{\Rd_u(T)} }{\F^{t-1} }},
\]
which is the same expression as the bound derived in Case 1.

\paragraph{Combining Cases 1 and 2.} Using the same bound for $L_t$ in
both cases, and exchanging the summation and expectation, we may
conclude by definition of $b_t$ that
\begin{align}\label{eqn:decaymidproof2}
\EE{\dotfrac{\Vd_u(T) + W(T)}{\Rd_u(T)}} &\leq
\EE{\dotfrac{W_0\decay^{T - \min\{\tau_1, T \}} + \sum_t
    \decay^{T-t}R_t u_t b_t}{\Rd_u(T)} }.
\end{align}
We now note that $b_t$ always equals $\alpha$, except for the very
first rejection at time $\tau_1$, in which case it equals $\alpha -
\frac{W_0}{u_{\tau_1}}$. Hence, we may write
\begin{align*}
\sum_t \decay^{T-t}R_t u_t b_t &= \sum_t \decay^{T-t}R_t u_t \alpha -
\decay^{T-\tau_1} W_0 \One{T \geq \tau_1} \\ &= \alpha \Rd_u(T) -
\decay^{T-\tau_1} W_0 \One{T \geq \tau_1}.
\end{align*}
Resubstituting this expression into bound \eqref{eqn:decaymidproof2}
yields
\begin{align*}
\EE{\dotfrac{\Vd_u(T) + W(T)}{\Rd_u(T)}} &\leq
\EE{\dotfrac{W_0\decay^{T - \min\{\tau_1, T \}} + \alpha \Rd_u(T) -
    W_0 \decay^{T-\tau_1 }\One{T \geq \tau_1} }{\Rd_u(T)} } \leq
\alpha,
\end{align*}
where the last inequality follows by verifying that it holds in the
three cases
\begin{align*}
  \{T < \tau_1 = \infty, \Rd_u(T) = 0\}, \quad \{T \geq \tau_1,
  \Rd_u(T) < 1\}, \quad \mbox{and} \quad \{T \geq \tau_1, \Rd_u(T)
  \geq 1\}
\end{align*}
separately. This completes the proof of the theorem.


\section{Summary}

In this paper, we make four main contributions---more powerful procedures
under independence, an alternate viewpoint of deriving online FDR procedures, 
incorporation of prior and penalty weights, and
introduction of a decaying-memory false discovery rate to handle
piggybacking and alpha-death. Numerical simulations in \appref{sims}
complement the theoretical results.  

\subsection*{Acknowledgments}
We thank A. Javanmard, R. F. Barber, K. Johnson, E. Katsevich,
W. Fithian and L. Lei for related discussions, and A. Javanmard for
sharing code to reproduce experiments in \citet{javanmard2016online}.
This material is based upon work supported in part by the Army
Research Office under grant number W911NF-17-1-0304, and
National Science Foundation grant NSF-DMS-1612948.


\bibliography{FDR}

\bibliographystyle{plainnat}






\end{document}